\documentclass{aa}
\usepackage[varg]{txfonts}
\usepackage{amsmath,amssymb}
\usepackage[colorlinks=true,linkcolor=blue,citecolor=blue,filecolor=blue,urlcolor=blue]{hyperref}
\usepackage{siunitx}
\usepackage{xcolor}
\DeclareGraphicsExtensions{.png,.pdf}

\DeclareMathOperator{\sgn}{sgn}

\begin{document}

\title{Close-in compact super-Earth systems emerging from resonant chains: slow destabilization by unseen remnants of formation}
\titlerunning{Slow Instabilities of Resonant Chains}
\authorrunning{M. Goldberg and A. Petit}
\author{Max Goldberg\inst{\ref{inst1}} 
\and Antoine C. Petit\inst{\ref{inst1}}}

\institute{
Laboratoire Lagrange, Université Côte d'Azur, CNRS, Observatoire de la Côte d'Azur, Boulevard de l'Observatoire, 06300 Nice Cedex 4, France\label{inst1}}
\date{}

\abstract 
{
Planet formation simulations consistently predict compact systems of numerous small planets in chains of mean motion resonances formed by planet-disk interaction, but transiting planet surveys have found most systems to be non-resonant and somewhat dynamically excited. A scenario in which nearly all of the primordial resonant chains undergo dynamical instabilities and collisions has previously been found to closely match many features of the observed planet sample. However, existing models have not been tested against new observations that show a steep decline in the resonant fraction as a function of stellar age on a timescale of $\sim 100$ Myr. We construct a simplified model incorporating Type I migration, growth from embryos, and N-body integrations continued to 500 Myr and use it to generate a synthetic planet population.
Nearly all systems exit the disk phase in a resonant configuration but begin slowly diffusing away from the center of the resonance. Dynamical instabilities can arise on timescales of tens or hundreds of Myr, especially when systems formed in disks with a convergent migration trap.
In this case, a secondary chain of smaller planets that remained at their birth location eventually breaks, destabilizing the inner resonant chain.
We also show that the instability statistics are well modeled by a Weibull distribution, and use this to extrapolate our population to Gyr ages. The close match of our modeled systems to the observed population implies that the high resonance fraction predicted by this class of models is in fact consistent with the data, and the previously-reported overabundance of resonant systems was a consequence of comparing simulations of early evolution to mature Gyr-old systems. This result also suggests that instabilities triggered by disk dissipation or other very early mechanisms are unlikely to be consistent with observed young systems.
}
\keywords{ Planets and satellites: formation --
             Planets and satellites: dynamical evolution and stability --
             Planet-disk interactions
           }

\maketitle

\section{Introduction}
Systems of multiple planets with sizes between Earth and Neptune orbiting with short orbital periods have been discovered in the hundreds by transit surveys and are now known to be one of the most common types of planetary systems in the galaxy \citep{Fressin2013,Petigura2013}. These \textit{Kepler}-like systems are increasingly understood to be a cohesive population with a set of correlated defining statistics, describing the physical properties of the planets and their orbital architectures \citep{Weiss2018,VanEylen2015,He2020}. Parameters such as planet multiplicity, orbital spacing, and typical planet sizes are well-understood for planets with radii between $1$ and $\rm 4~R_\oplus$ and periods $<100$ d. Formation models that aim to explain the origin of these systems must therefore focus on reproducing these statistics, rather than matching individual systems. 

Briefly, the typical \textit{Kepler}-like system can be summarized as follows: the planets are either super-Earths with radii between 1 and $\rm 1.6~R_\oplus$, or sub-Neptunes above $\rm 2~R_\oplus$, the latter of which have a large volatile envelope. At the time of formation these super-Earths and sub-Neptunes are a single population, with the radius difference emerging later as a consequence of core mass and insolation-dependent atmosphere loss \citep{Wu2019,Rogers2021,Owen2024}. In terms of orbital architecture, many systems have multiple planets that are closely spaced in period but not near mean-motion resonance, while other systems have just a single transiting planet \citep{Winn2015}. A small number of systems have many planets locked in pairwise mean motion resonance; these are the ``resonant chains.'' Producing this diversity of orbital architectures for small planet systems is a major challenge of planet formation theory.

A scenario that incorporates near-universal dynamical instabilities of primordial resonant chains is thus far the most successful model to reproduce the observed planetary census \citep{Pu2015,Izidoro2017}. In this model, resonant chains indeed form due to migration within the protoplanetary disk \citep{Cresswell2006,Cresswell2008} but are catastrophically disrupted after the dissipation of the disk. During the instability, planetary orbits intersect and planets collide with each other, the star, or are ejected from the system entirely. Eventually, the instability ends and leaves behind a more widely spaced and dynamically excited system. The period-ratio distribution, which was strongly peaked near MMR during the disk phase, becomes smooth and nearly uniform, a close match to the data \citep{Izidoro2017,Izidoro2021a}. The model also explains the observed eccentricity distributions, including the anomalously eccentric single-transiting systems \citep{VanEylen2019}, patterns of intra-system uniformity in planet mass and period ratio \citep{Goldberg2022,Lammers2023}, {and lower densities and smaller mutual inclinations of near-resonant planets \citep{Leleu2024}}.

A major limitation of formation models incorporating N-body integrations is that they are restricted by computational constraints to the early stages of system evolution. Many transiting planet systems have Gyr or even 10 Gyr ages, corresponding to $10^{10}-10^{12}$ orbits of the innermost planet, a daunting challenge for even the fastest single-threaded integrators. Although substantial evolution takes place within the time covered by these models, much dynamical evolution can occur later. It is therefore not clear whether formation models that extend only to $\sim 100~\rm Myr$ can be compared directly to data \citep{Emsenhuber2021}. 

A new, extremely powerful dataset is becoming available that promises to bridge this gap. The next generation of transit surveys, in particular \textit{TESS} \citep{Ricker2014}, have discovered 10--20 multiplanet systems around stars younger than $1$ Gyr, and including several younger than 100 Myr \citep{Hu2025}. While still a small minority of the known exoplanet population, these young and adolescent systems are a unique window into the formation environment of small planets, otherwise unobservable with current methods. In a population study, \cite{Dai2024} showed that in the youngest systems ($<100$ Myr), near-resonant configurations are ubiquitous, appearing in $>80\%$ of observed multiplanet systems. This fraction drops with a timescale of $\sim 100$ Myr to about $20\%$ for mature systems. Similar results have been corroborated by studies that derive stellar ages from galactic kinematics. \cite{Hamer2024} demonstrated that systems with planet pairs near resonance, particular second-order resonance, have smaller galactic velocity dispersions indicative of lower ages.

On its face, this result appears to be direct confirmation of the ``breaking-the-chains'' hypothesis, because that idea predicts that systems emerging from their natal disks should be in resonant configurations that later break. On closer look however, many details are unresolved. For one, it has frequently been suggested that an additional mechanism outside of the N-body dynamics is necessary to break enough chains to match the observed distribution. Reshuffling of the system during disk dispersal has been proposed as one candidate process \citep{Liu2017,Hansen2024,Pan2025}, but this likely occurs too early to explain the abundance of near-resonance for systems with ages $20-100$ Myr. In fact, it is not obvious what astrophysical process can occur over these timescales.

In this work, we demonstrate that resonant chains can unravel on timescales that match the observed decline in near-resonant frequency purely through N-body dynamics. In our model, planets grow out in a narrow ring \citep{Batygin2023a,Ogihara2024} and then through Type I migration form resonant chains that resemble observed resonant systems. However, in most of them, chaotic diffusion gradually drives the system away from a fixed point until the resonant protection is lost and the planets undergo a dynamical instability. Importantly, the rate of this diffusion is frequently very slow, with a long tail, reproducing the observed frequency of near-resonance. Previous work has demonstrated that systems that begin resonant but later went unstable can readily match the transiting planet population \citep{Izidoro2017,Izidoro2021a,Goldberg2022}. Our primary goal in this work is to begin to understand how, why, and when this instability happens.

Section \ref{sec:methods} describes our simulation setup and computational methods, and the results of these simulations in relation to the observed exoplanet population are presented in Section \ref{sec:results}. We explore the details of the dynamics in Section \ref{sec:discussion} and describe the limitations of our results. Section \ref{sec:conclusion} summarizes our findings.

\section{Methods}
\label{sec:methods}
We performed a suite of N-body simulations accounting for the growth of planets in the gaseous disk phase, planet-disk interactions, removal of the disk, and post-disk dynamical evolution of the system. 

All simulations used the N-body code \texttt{rebound} with the \textsc{trace} integrator \citep{Lu2024}. \textsc{trace} is a hybrid time-reversible integrator that switches from the classical Wisdom-Holman Hamiltonian splitting \citep{Rein2015} to direct Bulirsch-Stoer (BS) integration during close encounters. In comparison to the widely-used MERCURY code \citep{Chambers1999} (and the similar \texttt{rebound} implementation \textsc{mercurius} \citep{Rein2019}), \textsc{trace} is uniquely well-suited to this problem for several reasons. First, the switching distance is dynamically computed at each timestep rather than at the beginning of the integration, as it is in \textsc{mercurius}. While technically breaking symplecticity, this choice guarantees that the switching criteria remains robust even as planets grow or migrate \citep{Hernandez2023,Lu2024}. Secondly, \textsc{trace} properly resolves close pericenter approaches for highly-eccentric orbits, which can arise during dynamical instabilities. Finally, during close encounters, \textsc{trace} uses a much faster BS integrator than the IAS15 integrator used by \textsc{mercurius}, resulting in a major speed-up for very violent systems dominated by close encounters without a meaningful loss of accuracy. These benefits allowed us to perform significantly longer integrations and uncover slower instabilities.

The initial conditions of the simulations are summarized in Table~\ref{tab:init}. Following \cite{Batygin2023a}, who argue that planetary growth should commence from rings in which solids have concentrated, we initialize simulations at time $t=0$ by placing embryos of mass $\rm 0.2~M_\oplus$ with semi-major axes drawn from a Gaussian distribution with mean 1 au and standard deviation $\sigma$ around a central star of mass $\rm 1~M_\odot$ and radius $\rm 1~R_\odot$. Their eccentricities and inclinations were drawn from a Rayleigh distribution with mode $10^{-3}$ and $0.5\times 10^{-3}$, respectively, and their longitudes of pericenter, longitudes of ascending node, and mean longitudes were drawn from a uniform distribution from $[0,2\pi)$. All particles were set to be fully active. The timestep was set at the beginning of the integration to be $1/30\times$ the period of the innermost embryo, and periodically adjusted if it decreased below $1/12\times$ the period of a body due to inward migration. All bodies other than the star were set to have a physical radius corresponding to a density of $\rm 3~g ~cm^{-3}$ and, if they approached closer than the sum of their radii, underwent perfect mergers conserving total mass and linear momentum. 

\begin{table}[h]
    \caption{Initial conditions of simulations}
    \centering
    \begin{tabular}{c|c|c|c|c}
         Run name & $M_\textrm{tot} \ (\rm M_\oplus)$ & $\sigma$ (au) & $\alpha_\text{in}$ & \# of sims \\   
         \hline
         \texttt{20flat} & 20 & 0.2 & 0.0 & 40 \\
         \texttt{40steep} & 40 & 0.2 & -0.5 & 40 \\
         \texttt{10straight} & 10 & 0.1 & 1.5 & 40 \\
         \texttt{20straight} & 20 & 0.2 & 1.5 & 40 \\
         \texttt{40straight} & 40 & 0.2 & 1.5 & 40
    \end{tabular}
    \label{tab:init}
\end{table}

\begin{figure}
    \centering
    \includegraphics[width=\linewidth]{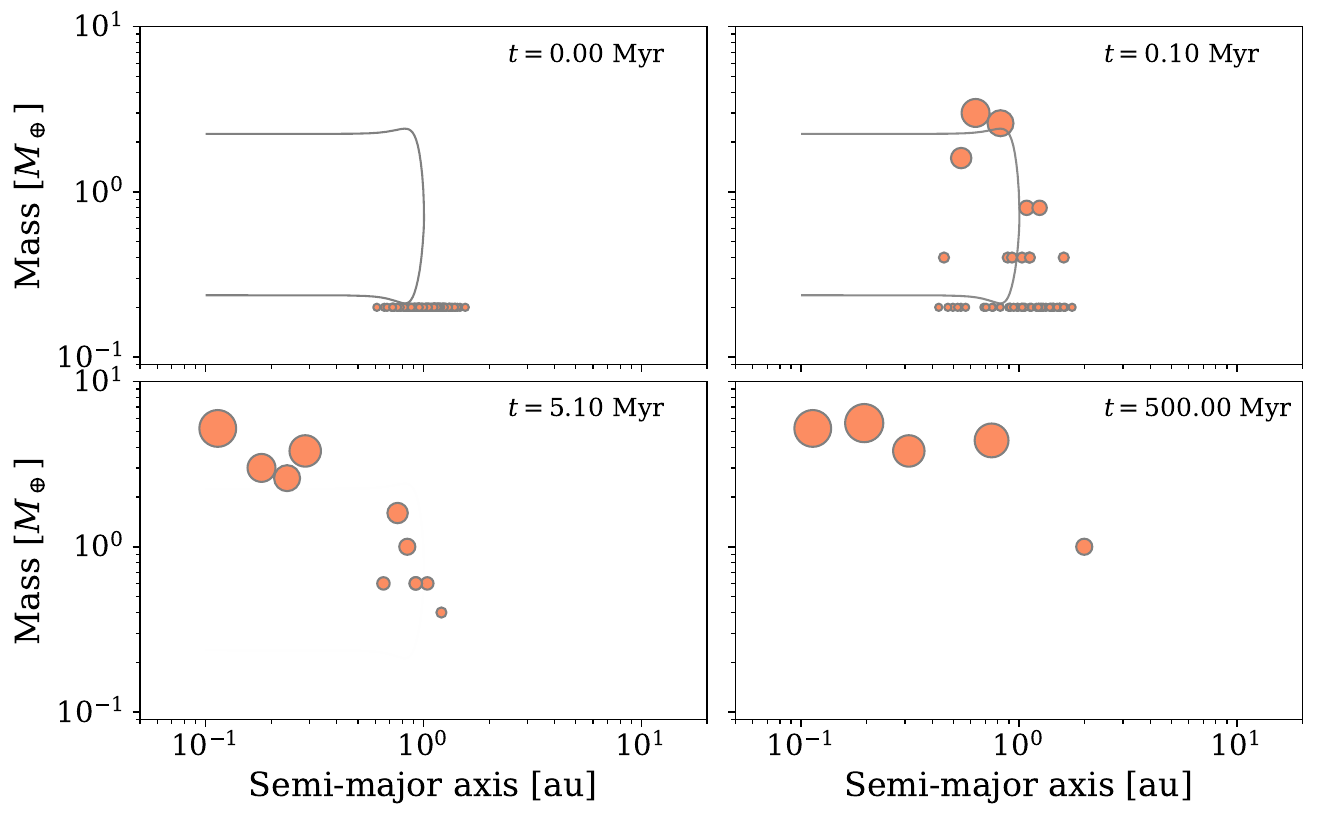}
    \caption{Evolution of a typical simulation from the \texttt{20flat} suite, which includes a migration trap at 1 au. Embryos accrete from a ring (top left panel) and grow until they saturate the corotation torque (top right, zero torque locations in gray), then migrate inwards in resonant convoys. At the end of the disk phase (bottom left), there is an inner resonant chain and an outer, less massive, and more compact one. By the end of the simulation (bottom right), the system has experienced a dynamical instability, breaking the resonances and merging planets.}
    \label{fig:exsim}
\end{figure}

We included the effects of Type I migration and damping on the embryos. The gas disk had a surface density of
\begin{equation}
    \Sigma(r,t) = \Sigma_0 \exp(-t/\tau_\textrm{diss}) \left(\frac{r}{r_0}\right)^{-p}
    \label{eq:disk}
\end{equation}
where the power law disk slope is
\begin{equation}
    p=  \frac{1}{2} \left(\alpha_\textrm{out} + \alpha_\textrm{in}) + (\alpha_\textrm{out} - \alpha_\textrm{in})  \tanh\left(\frac{r-r_0}{w}\right)\right),
\end{equation}
$\Sigma_0 = \rm 1600~g~cm^{-2}$, $\tau_\textrm{diss}=\rm 1~Myr$, $\alpha_\textrm{out}=3/2$, $r_0=1\rm~au$, $w=0.15\rm~au$, and $\alpha_\textrm{in}$ differs by simulation (Table~\ref{tab:init}). A constant aspect ratio was set to $h/r=0.05$. This parametrization represents a disk with a pressure maximum at 1~au, and a transition from a 3/2 power-law profile in the outer disk to a typically flatter profile in the inner disk. Such a profile is frequently found in simulations of disk evolution with magnetic winds that efficiently carve the inner disk \citep{Suzuki2016,Kunitomo2020}. The disk was dissipated exponentially on a timescale of 1~Myr, and then at $t=\rm 5~Myr$ the disk was rapidly removed on a timescale of $10^4$ yr to represent disk photoevaporation. At $t=\rm 5.1~Myr$ the disk and its corresponding forces were removed completely. Each system was then integrated to $t=\rm 500~Myr$ under purely Newtonian gravity. The evolution of a typical simulation is shown in Fig.~\ref{fig:exsim}.

Embryos felt type I migration forces corresponding to Lindblad and corotation torques, computed using the equations of \cite{Paardekooper2010,Paardekooper2011}. These formulae incorporate a dependence on the local power law index of surface density and temperature and account for saturation of the corotation torque, for which we use a Shakura-Sunyaev turbulent $\alpha=10^{-4}$. We also included eccentricity and inclination damping forces according to the prescription of \cite{Cresswell2006,Cresswell2008}, which account for reduction in damping at high inclination and eccentricity. For disks with $\alpha_\textrm{in} < \alpha_\textrm{out}$, a migration trap at 1 au appears for a specific range of planet masses. The trap can hold planets in the outer system until they grow large enough to saturate the corotation torque, and quickly migrate inwards. While not derived from a specific physical model, this simulation setup is intended to represent a disk with a broad gas pressure bump and dust trap that triggers planetesimal and embryo formation \citep{Flock2017}.

\begin{figure}
    \centering
    \includegraphics[width=\linewidth]{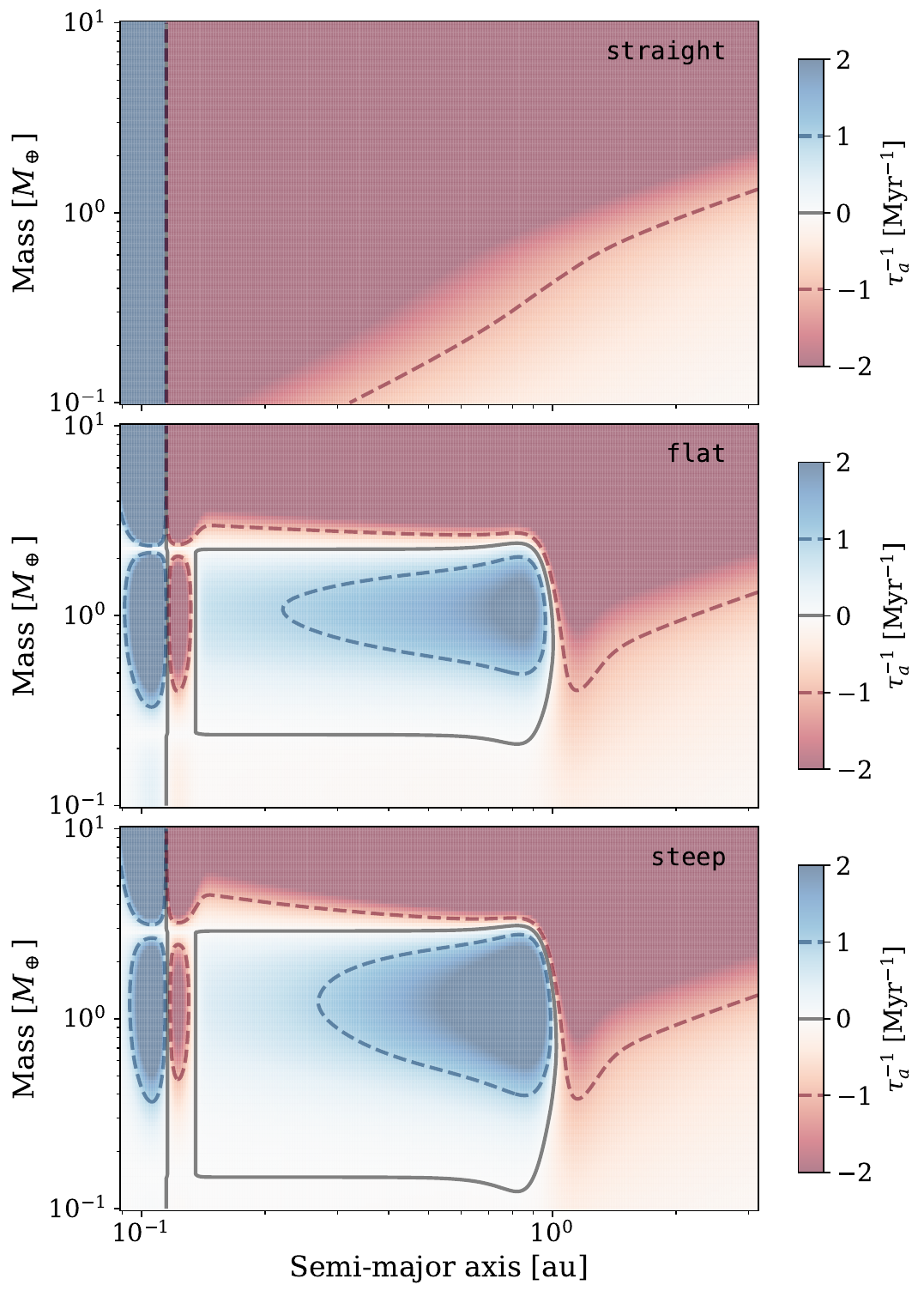}
    \caption{Migration map of the three disk profiles used (Eq.~\ref{eq:disk}) at $t=0$ including the repulsive inner edge. Positive values of $\tau_a^{-1}$ indicate outward migration. The solid gray line marks the zero-torque locus and dashed lines show the locations where $\tau_a = \pm 1$ Myr. For the flat and steep disks, there is a migration trap near the pressure bump at 1 au that suppresses migration until planets reach $\sim \rm 2~M_\oplus$, at which point they saturate the corotation torque and migrate inwards.}
    \label{fig:migmap}
\end{figure}

The disk is also believed to be truncated at short radii when magnetic forces from the star are more important than accretion ram pressure. These sharp decreases in surface density also act as planet traps where inward migration is stopped and strong outward torques are possible \citep{Masset2006}. While some previous work prescribed a falloff in surface density at the inner edge and let corotation torques provide outward migration \citep[e.g.,][]{Izidoro2017}, \cite{Ataiee2021} showed that the standard torque prescription based on local power laws fails in these regimes. In fact, they found the trap to be much stronger than predicted, so that one planet at the inner edge is capable of halting the migration a resonant chain of several planets. We use a slightly modified version of their recommended prescription, in which we multiply the local migration rate $\tau_a=\dot{a}/a$ by a coefficient
\begin{equation}
        f_\textrm{trans} = 1 + \mathcal{A}(\tanh\mathcal{R}-1) - \sgn(\tau_a) \mathcal{B} \cdot (\mathcal{R} - \mathcal{C})\exp(-\mathcal{R}^2)
\end{equation}
where $\mathcal{R} = (r - r_\textrm{in})/\mathcal{W}$, $\mathcal{A} = 0.45$, $\mathcal{B} = 20$, $\mathcal{C} = 0.226$, $\mathcal{W}=0.012$ au, and $r_\textrm{in}=\rm 0.113~au$. This leads to a migration trap very close to 0.1~au, which is in very close agreement with the inner edge of \textit{Kepler} systems being clustered near 10 d \citep{Mulders2018}. Figure~\ref{fig:migmap} shows a migration map at $t=0$ of this disk with the three values of $\alpha_\textrm{in}$ used in this work.

\section{Results}
\label{sec:results}
The vast majority of our simulation runs proceed in a similar fashion. Briefly, the embryos merge with each other near their starting location until they are massive enough to efficiently migrate inwards. Pairs frequently migrate convergently, either because the outer embryo is more massive than the inner or because the inner one is trapped at the inner disk edge. Embryos consistently capture into chains of mean motion resonances, most commonly first-order ones, although higher-order and co-orbital resonances also appear. While dynamical instabilities are common in the disk phase, damping and migration rapidly recapture the systems into resonance and in nearly all of our simulation runs, the systems exit the disk phase in a mean-motion resonant configuration. 

Following the disk dissipation, the system evolution takes one of two distinct paths. In the quiescent pathway, the resonant chain remains effectively unchanged over the entirety of the simulation, with purely periodic changes in orbital elements from resonant and secular evolution, reaching $t=\rm 500~Myr$ with librating resonant angles. On the other hand, many systems experience a dynamical instability sometime between $t=\rm 5.1~Myr$ and $\rm 500~Myr$ where the resonances become broken and the system is rapidly chaotic. In most cases, planetary orbits cross and planets collide very soon after the resonant angles begin circulating. The remaining systems are non-resonant, more widely spaced, more eccentric and inclined, and the planets are fewer and more massive. The quiescent pathway is interpreted as producing the rarely-observed resonant chains such as TRAPPIST-1 and TOI-178 \citep{Gillon2017,Leleu2021}, whereas the unstable track is responsible for the bulk of the known population of small planets in compact systems, which are non-resonant. The ability of a single process to reproduce both populations is the essence of the ``breaking-the-chains'' hypothesis originally proposed by \cite{Izidoro2017}.

\subsection{Planet accretion, migration, and resonance capture}
The initial phase of planet accretion proceeds very quickly. Embryos quickly excite each other to crossing orbits and grow to several Earth masses. Typically, only 10--20 embryos remain by $t=\rm 100~kyr$. In models with a pressure bump, embryos that reach the minimum mass to saturate the corotation torque migrate inwards quickly, sometimes resonantly entraining smaller planets that remain interior to them. Migration occurs earlier in \texttt{straight} disks, with planet masses set by a balance of local growth rates and the timescale to exit the ring \citep{Batygin2023a}. Migration always terminates at the inner edge located at 0.1 au, where the powerful planet trap prevents the chain from falling further towards the star.

The relative rates of migration for adjacent planets depends on their masses and the local disk profile, but convergent migration is frequently achieved because the inner planet is trapped by the inner edge directly or indirectly. Under these conditions, adjacent planet pairs capture efficiently into mean motion resonances, gradually building a large resonant chain anchored at the inner edge. Dynamical instabilities are repeatedly triggered during the disk phase either because migration dynamically overpacks the system or because of resonant overstability \citep{Goldreich2014}. However, high gas densities in the inner disk allow the broken chain to quickly reform. By the end of the disk phase at $t=\rm 5.1~Myr$, 90--100\% of systems are clearly resonant chains. The most common resonances are 5:4, 4:3, and 3:2 for the straight simulations and 4:3, 3:2, and 2:1 for \texttt{20flat} and \texttt{40steep}. Perhaps the biggest difference between the simulations is that those with a pressure bump frequently contain adjacent pairs of planets that are non-resonant: approximately 10\% of pairs in \texttt{20flat} and \texttt{40steep} are further than 5\% from a first- or second-order mean motion resonance, compared to $< 1\%$ in the \texttt{straight} simulations. Additionally, the coupling between mass and migration speed leads to a more pronounced mass ordering in the \texttt{straight} simulations so that the innermost planets are almost always the most massive. In contrast, in simulations with a migration trap, all planets interior to 0.5 au tend to have a similar mass.

\subsection{Population statistics}
\begin{figure}
    \centering
    \includegraphics[width=\linewidth]{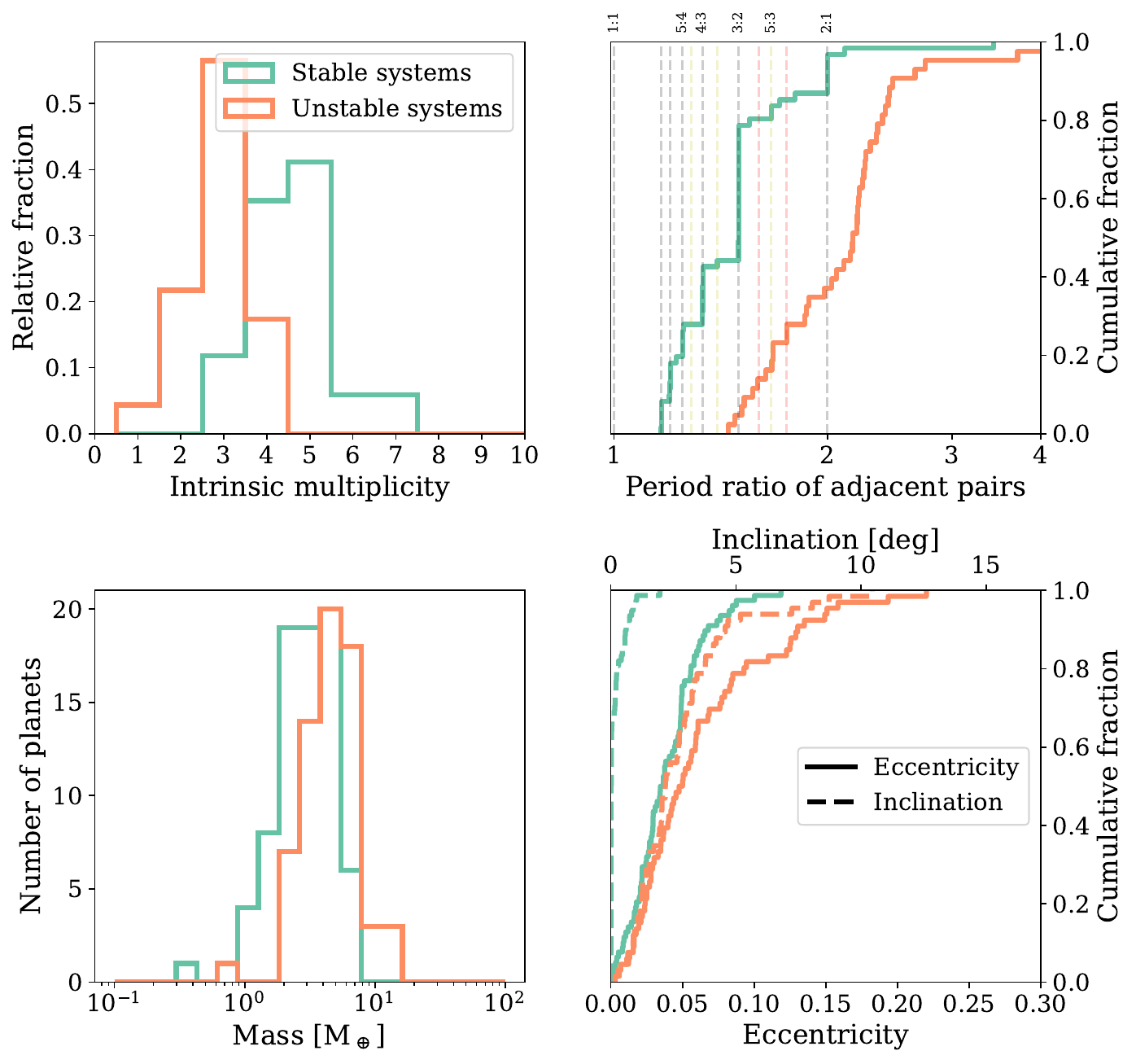}
    \caption{Summary population statistics for the 40 synthetic planetary systems in the simulation suite \texttt{20flat} at the end of the $\rm 500~Myr$ integration. Orange and cyan curves show planets in systems that did and did not undergo a dynamical instability after the dissipation of the protoplanetary disk, respectively. Only planets with orbital periods $<\rm 100~d$ are included.}
    \label{fig:20flatpopstats}
\end{figure}

Figure \ref{fig:20flatpopstats} summarizes the statistics of the \texttt{20flat} synthetic population at the end of the 500~Myr integrations, divided by whether the system experienced an instability after the end of the disk phase. Similar figures for the other simulation suites are presented in Appendix~\ref{app:pop}. In these figures, all planets within 100 d have been included.

Some features are common to all simulation sets. Stable systems have more planets of lower mass, compared to unstable systems, and the period ratios of adjacent planets are dominated by resonant ratios. Most of these systems are extremely flat because their inclinations were damped by the protoplanetary disk and first-order resonances do not involve inclination, although a few have inclinations of $\sim 1^\circ$. Eccentricities have been moderately excited by resonances to $\sim 0.05$, a quantity which is set by the planet masses and balance of resonant excitation during migration vs. eccentricity damping in the disk \citep{Terquem2019}.

Unstable systems have qualitatively different features. In the closely-packed small planet regime, systems dynamically relax by way of collisions and mergers, thus reducing the number of planets but increasing the mass of the remaining ones. Prior resonances are systematically broken by the powerful planet-planet interactions during orbit crossing. The only source of energy dissipation is through inelastic mergers, which are insufficient to reform mean motion resonances. Final period ratios are determined by dynamical stability and dependent on planet masses and multiplicity \citep{Pu2015}. High mass runs thus have more widely spaced systems, from a median period ratio of 1.5 in \texttt{10straight} to 2.49 in \texttt{40straight}. Instabilities also leave eccentricities and inclinations in an excited state correlated with planet mass, although inelastic collisions sometimes leave systems anomalously flat or circular \citep{Esteves2020}.

The fraction of systems that went unstable before the end of the 500~Myr integrations differs dramatically between simulation suites, from a low of 32.5\% in \texttt{10straight} to a high of 92.5\% in \texttt{40steep}. Two patterns are clear from Table~\ref{tab:results}. First, for the same disk profile, simulations with higher $M_\textrm{tot}$ have a higher instability fraction. This is expected because over the regime of interest, the final resonance that planets capture into during migration depends much more strongly on the local gas surface density than on the planet masses \citep{Kajtazi2023}. High mass planets therefore capture into the same resonances as low mass planets, but are much more closely packed dynamically \citep{Petit2020a,Goldberg2022a}. The second clear trend is that, for the same $M_\textrm{tot}$, systems that formed in disks with a migration trap have a higher instability fraction. For example, 57.5\% of systems in \texttt{20flat} went unstable, compared to 35\% in \texttt{20straight}. Likewise, \texttt{40steep} had a 92.5\% unstable fraction compared to 80\% in \texttt{40straight}. We explore the possible origins of this trend in Section~\ref{sec:unstabphysics}.

\begin{table*}
    \caption{Summary of simulation results}
    \centering
    \begin{tabular}{c|c|c|c|c}
         Run name & Unstable fraction (\%) at $\rm 500~Myr$ & Extrapolated fraction unstable (\%) at $\rm 5~Gyr$ & Weibull $\mathcal{T}$ (Myr) & Weibull $k$ \\   
         \hline
         \texttt{20flat} & 57.5 & $87.6^{+7.2}_{-6.7}$ & $648\pm 393$ & $0.37\pm 0.07$ \\
         \texttt{40steep} & 92.5 & 100.0 & $23 \pm 7$ & $0.38\pm 0.05$ \\
         \texttt{10straight} & 32.5 & $59.0^{+0.7}_{-0.8}$ & $6971\pm 106$ & $0.35\pm 0.05$ \\
         \texttt{20straight} & 35.0 & $77.1^{+8.0}_{-5.7}$ & $2341\pm 706$ & $0.53\pm 0.11$ \\
         \texttt{40straight} & 80.0 & $98.4^{+1.2}_{-2.0}$ & $111\pm 50$ & $0.37\pm 0.05$
    \end{tabular}
    \label{tab:results}
\end{table*}

\subsection{Timing of the instability}
\begin{figure}
    \centering
    \includegraphics[width=\linewidth]{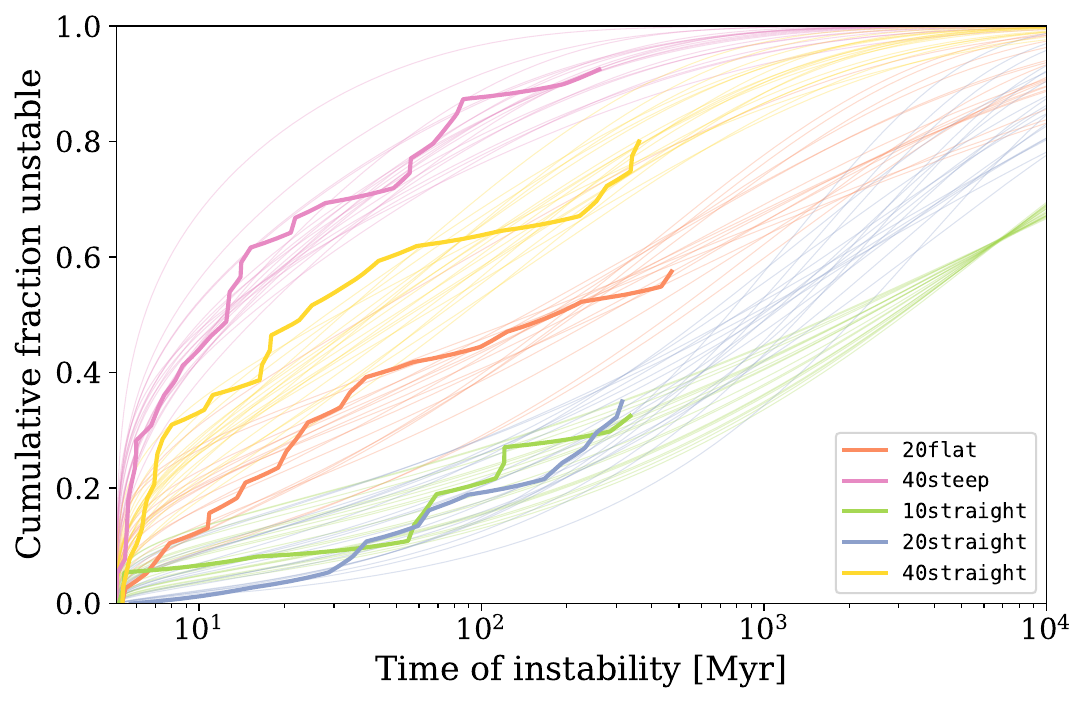}
    \caption{The distribution for the post-disk instability time in each simulation suite (solid lines). Thin lines are samples from the fits to a Weibull distribution, widely used in failure rate statistics.}
    \label{fig:insttime}
\end{figure}

The time required for a resonant system to go unstable, and thus transform into a non-resonant system, is of particular importance thanks to recent results of the population characteristics of young systems and an increasing number of clear trends in planetary system architectures with age. Our goal in particular is to determine whether purely N-body dynamics can reproduce the rapid decline in the fraction of systems observed near mean motion resonance on a timescale of approximately $\rm 100~Myr$ \citep{Dai2024}. Figure~\ref{fig:insttime} shows the cumulative distributions for the time of instability of our systems, separated by simulation suite. The ``instability'' is defined as the first time that a planet with period $<100$ d at the end of the disk phase had its mass or semi-major axis change by more than 5\%, and thus does not include instabilities that occur purely in the outer system that do not propagate to the inner planets. 

We found that the instability time distribution is well-modeled by a Weibull distribution \citep{Weibull1951}, the cumulative function of which is given by
\begin{equation}
    \mathcal{W}(t_\textrm{inst}) = 1 - \exp\left(-\left(\frac{t_\textrm{inst}-t_0}{\mathcal{T}}\right)^k\right).
\end{equation}
Here, $t_0$ represents the earliest possible time for an instability, defined to be the disk dissipation time $t_0=\rm 5.1~Myr$, $\mathcal{T}$ is a scale parameter, and $k$ is a shape parameter. While not derived from a specific physical model, the Weibull distribution is widely used in failure rate statistics to describe the time of failure of a complex system made of multiple components\footnote{Indeed, Weibull's original motivation for this distribution was to study the lifetime of a chain that breaks upon the failure of a single link.} and its parameters have a straightforward interpretation. The scale parameter $\mathcal{T}$ is defined simply as the time at which a $1/e$ fraction of systems have failed. Furthermore, in this context, one speaks of the hazard function, $h(t)=-\mathcal{W}'(t)/(1-\mathcal{W}(t))$ which describes the instantaneous rate of failure for systems still functioning at time $t$. For the Weibull distribution, the hazard function takes the form
\begin{equation}
    h(t_\textrm{inst}) = \frac{k}{\mathcal{T}}\left(\frac{t_\textrm{inst}-t_0}{\mathcal{T}}\right)^{k-1}
\end{equation}
which is decreasing in time for $k<1$ and increasing for $k>1$ (for $k=1$ the Weibull distribution simplifies to the exponential distribution). Essentially, $k<1$ represents a sort of ``infant mortality'' wherein many systems fail early, but older ones are increasingly robust and have a reduced likelihood of failure.

We fit the simulated instability time distribution of each simulation suite to a Weibull distribution by maximizing the total likelihood. Systems that remained stable over the duration of the integration were assigned likelihood $1-\mathcal{W}(\rm 500~Myr)$. Uncertainties in the distribution parameters and sample draws from the fit were computed by treating the estimated Hessian matrix returned by the BFGS optimizer implemented in \texttt{scipy} as a covariance matrix \citep{Nocedal2006}. 

The Weibull parameters best fits and uncertainties are given in Table~\ref{tab:results}. Draws from the fit are also shown in Fig.~\ref{fig:insttime} along with the simulation-derived empirical distribution function. The instability timescale $\mathcal{T}$ spans several orders of magnitude across simulation suites. There is a clear dependence on total mass that is steeper from $M_\textrm{tot}=\rm 20~M_\oplus$ to $\rm 40~M_\oplus$ than from $\rm 10~M_\oplus$ to $\rm 20~M_\oplus$, suggesting that these resonant chains span a stability limit driven by the overlap of resonances \citep{Quillen2011,Petit2020a}. The simulations with a pressure bump also have $\mathcal{T}$ that is a factor of $\sim 5$ times smaller than the corresponding ones without.

In contrast, the shape parameter $k$ is exceptionally consistent across our simulations. Indeed, four of the five suites are within one sigma of $k=0.37$ and \texttt{20straight} is 1.5 sigma away from that value. The exact interpretation of this uniformity is not clear, but it suggests that an underlying mechanism sets the instability time across all of our simulations and differs only in its strength.

\subsection{Simulated observations}
\label{sec:synthobs}
In order to directly compare our results to the observed transiting planet sample, we applied a synthetic observation algorithm to determine which planetary systems would be observed by a transiting survey. To do this, for each simulated system, we simulate observations from 1000 random viewing angles and record which planets transit the star using the CORBITS package \citep{Brakensiek2016}, constructing a large synthetic observation catalog.

Our integrations only continued to 500~Myr due to computation time constraints, but typical \textit{Kepler} systems are much older. We take a fiducial age of $\sim \rm 5~Gyr$. As resonant chains continued going unstable until the end of our simulations, we expect that even fewer will be resonant at late times. We extrapolated the instability fraction to 5~Gyr by sampling the fits to the Weibull distribution and computing the cumulative value at 5~Gyr, taking the median and $1\sigma$ ranges for each simulation suite. Then, we created a new observed catalog by resampling the previous one according to the new unstable fraction. For example, 57.5\% of the \texttt{20flat} simulations went unstable by 500~Myr, but extrapolating the Weibull fit leads to $87.6^{+7.2}_{-6.7}\%$ unstable systems at 5~Gyr. We generate a new observed catalog where 87.6\% of systems are drawn from unstable systems and the remainder from the stable systems, while still sampling random viewing angles. Previous work performed similar resampling \citep{Izidoro2017,Izidoro2021a}, but only in an ad hoc way to match the observed exoplanet demographics rather than deriving it from the instability statistics.

The synthetic observations are particularly important because the stable and unstable systems are observed in different ways. Stable systems are very flat, leading to typically an all-or-nothing discovery outcome depending on their sky-plane inclination. We therefore expect observed resonant systems to be mostly representative of their true configuration, at least in the inner system. In contrast, unstable systems have moderate inclination, so it is uncommon to observe the entire system in transit. Frequently, only a single planet is observed, or a planet is missed in between two observed ones, biasing both the observed multiplicity and period ratio distribution. 

\begin{figure}
    \centering
    \includegraphics[width=\linewidth]{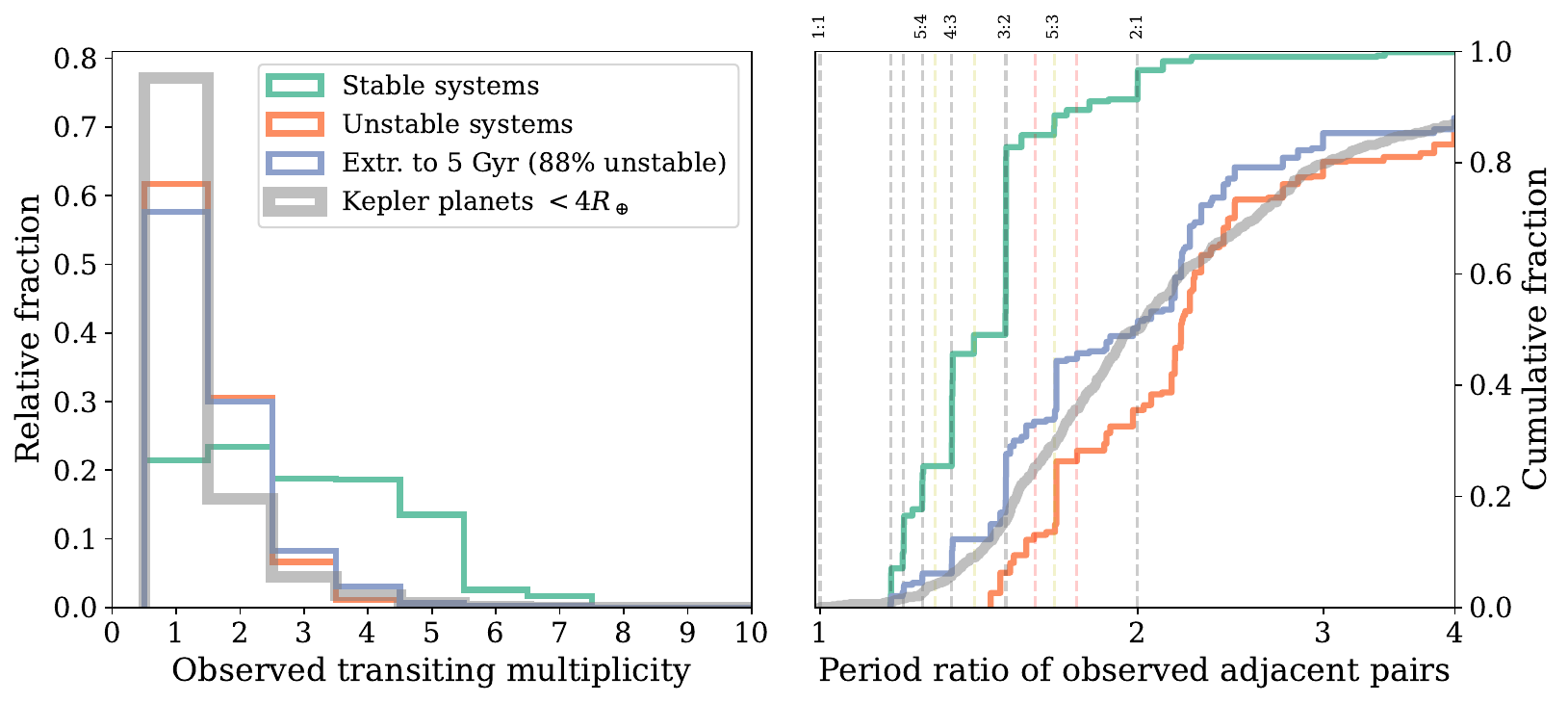}
    \caption{The distributions of transiting multiplicity and, for multitransiting systems, period ratio of adjacent planets, for the \texttt{20flat} simulation suite. For each system, 1000 ``observations'' are taken along different lines-of-sight to determine which planets would be observed. The purple line is a set of planetary systems resampled from the original \texttt{20flat} set such that 88\% of them had a post-disk instability, corresponding to the extrapolated value of the Weibull distribution at $t=\rm 5~Gyr$ (see text for details). Light gray lines are the observed \textit{Kepler} population for systems where all transiting planets were smaller than $\rm 4~R_\oplus$.}
    \label{fig:20flatobspopstats}
\end{figure}

\begin{figure}
    \centering
    \includegraphics[width=\linewidth]{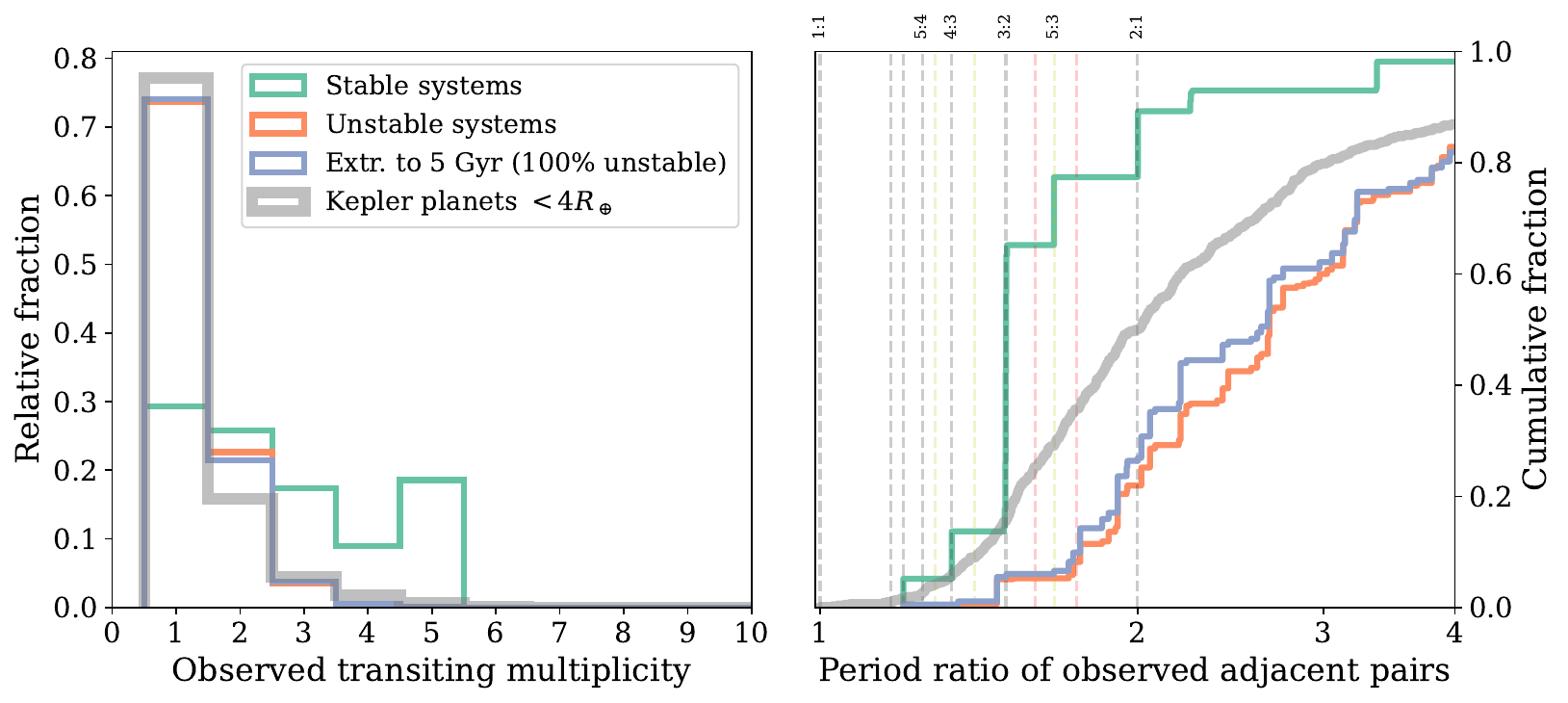}
    \caption{Same as Fig.~\ref{fig:20flatobspopstats} for run \texttt{40steep}.}
    \label{fig:40steepobspopstats}
\end{figure}

\begin{figure}
    \centering
    \includegraphics[width=\linewidth]{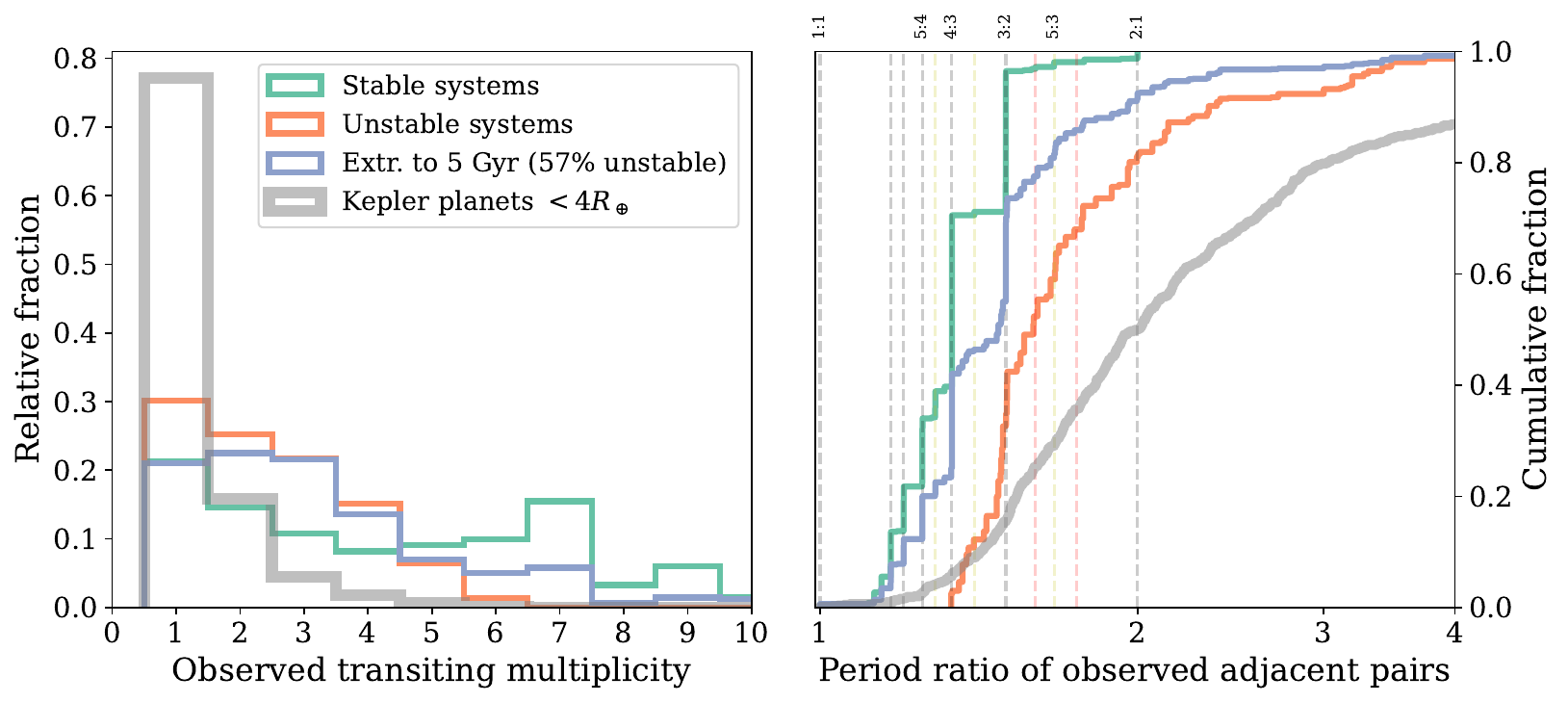}
    \caption{Same as Fig.~\ref{fig:20flatobspopstats} for run \texttt{10straight}.}
    \label{fig:10straightobspopstats}
\end{figure}

\begin{figure}
    \centering
    \includegraphics[width=\linewidth]{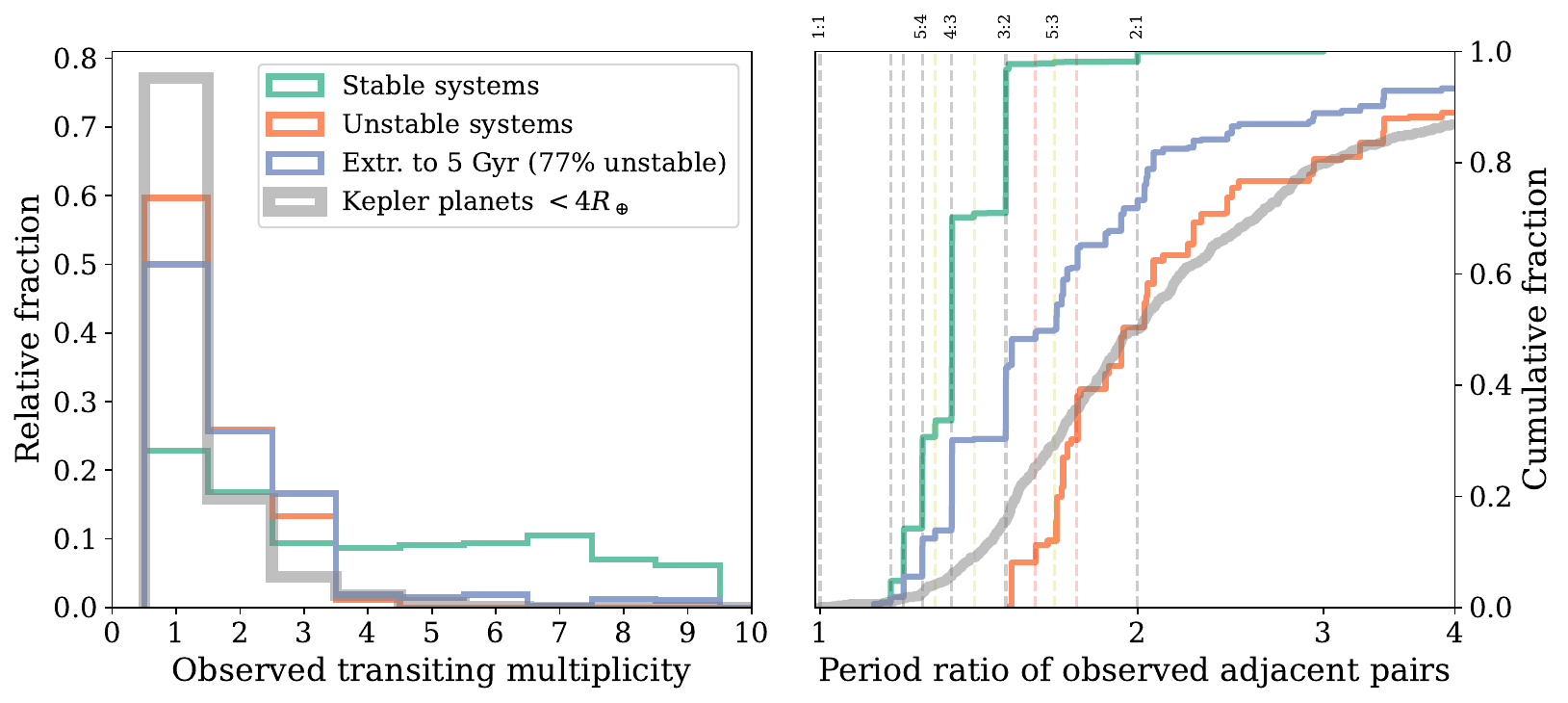}
    \caption{Same as Fig.~\ref{fig:20flatobspopstats} for run \texttt{20straight}.}
    \label{fig:20straightobspopstats}
\end{figure}

\begin{figure}
    \centering
    \includegraphics[width=\linewidth]{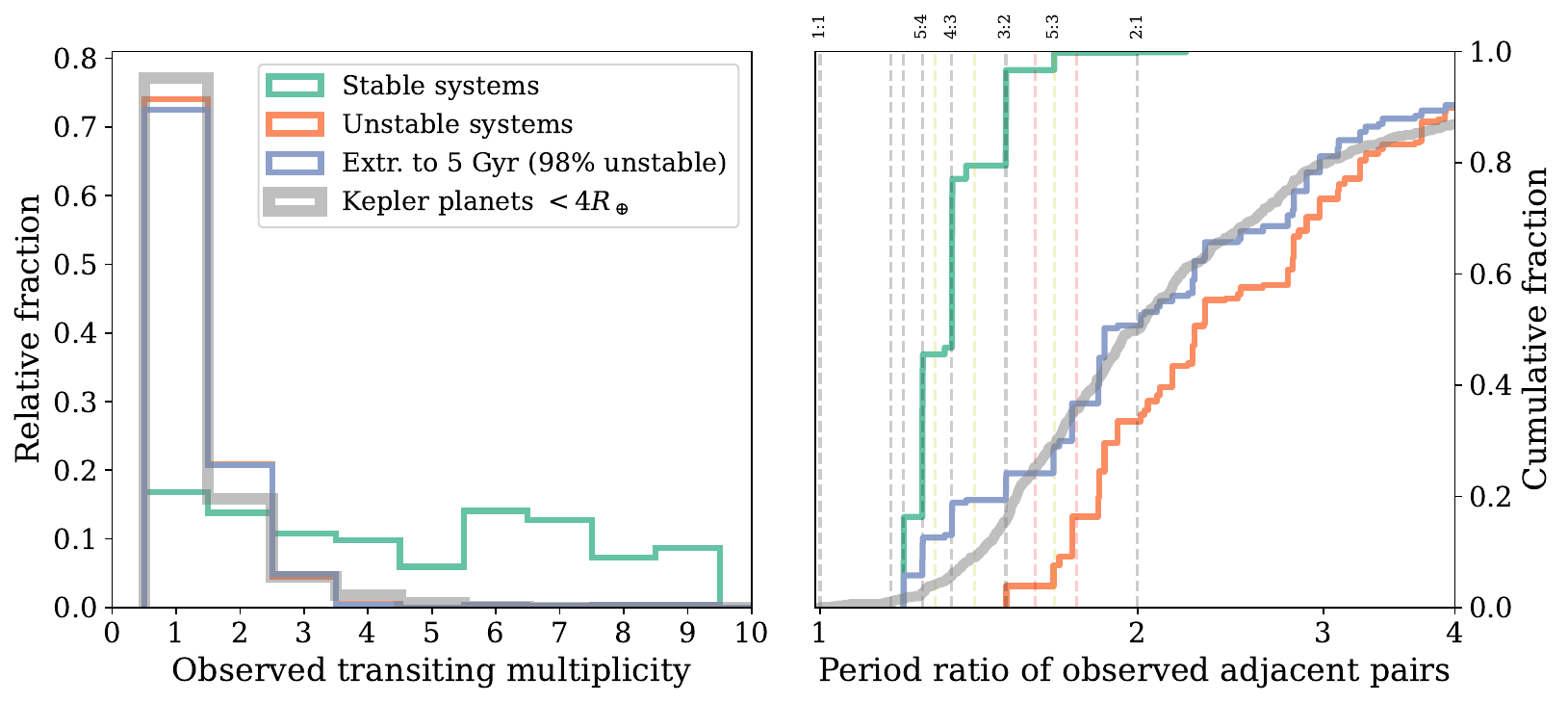}
    \caption{Same as Fig.~\ref{fig:20flatobspopstats} for run \texttt{40straight}.}
    \label{fig:40straightobspopstats}
\end{figure}

Figures~\ref{fig:20flatobspopstats}--\ref{fig:40straightobspopstats} show the inferred multiplicity and period ratio statistics that would be observed by a transiting survey, split into the stable and unstable systems and the resampled catalog in purple. We also plot the multiplicity and period ratio distributions from candidate and confirmed planets detected by the \textit{Kepler} mission, taken from the Exoplanet Archive.\footnote{\url{https://exoplanetarchive.ipac.caltech.edu/}} To compare directly with simulations, only planets with periods $<\rm 100~d$ are included and systems with a giant planet (defined as radius $>\rm 4~R_\oplus$) are excluded. 

As seen in previous studies, planetary systems that underwent a dynamical instability closely resemble the observed transiting planet population. The high mutual inclinations from unstable systems leads to a large number of systems with only a single planet in transit, explaining the so-called \textit{Kepler} dichotomy between single and multi-transiting systems \citep{Johansen2012}. Furthermore, the observed period ratio distribution depends strongly on the total mass, but matches the observed period ratio distribution. 

The strength of the produced dichotomy depends strongly on the initial mass. Higher mass planets systems generate larger mutual inclinations during the instability, which directly strengthens the dichotomy \citep{Izidoro2017,Millholland2021}. Furthermore, in our simulations, low mass systems have a smaller instability fraction and so the final system catalog is dominated by high multiplicity resonant chains. We find that the \texttt{40steep} and \texttt{40straight} simulations most closely reproduce the observed \textit{Kepler} multiplicity distribution of all of the tested configurations. The lower mass \texttt{20flat} and \texttt{20straight} simulations still produce a dichotomy, although it is too weak to match the nearly $80\%$ of single transiting systems observed by \textit{Kepler}. Finally, the lowest mass \texttt{10straight} simulation, with its low inclinations and low rate of instability, does not show a distinct dichotomy.

The orbital spacing of planetary systems post-instability is much broader for more massive systems. The observed \textit{Kepler} period ratio distribution is most closely matched by the unstable systems that started with $\rm 20~M_\oplus$ (orange curve, Figs.~\ref{fig:20flatobspopstats} and \ref{fig:20straightobspopstats}). However, the overall period ratio distribution is strongly influenced by the stable systems because, despite their rarity, they are frequently observed as multiples. In the end, a very close match to the observed distribution is obtained in the \texttt{40straight} run, although \texttt{20flat} is a satisfactory match.

Now that we have confirmed that the final states of these simulations are an acceptable match to observed mature systems, we consider the new constraint of the time evolution of the fraction of systems near mean-motion resonance. In the absence of robust dynamical information for most systems, \cite{Dai2024} defined ``near-resonant'' systems as having a planet pair with $-0.015\leq \Delta\leq 0.03$ for first-order resonances and $-0.015\leq\Delta\leq0.015$ for second-order resonances ($\Delta$ is defined in Section~\ref{sec:reschains}). This period ratio-based definition is different than our stable/unstable distinction because some instabilities can occur without breaking all resonances, and because unstable systems occasionally produce pairs that are serendipitously near MMR in period ratio. For the latter reason, even a catalog of entirely non-resonant systems will be measured as having a 10--20\% incidence of resonance by the empirical criterion.

\begin{figure}
    \centering
    \includegraphics[width=\linewidth]{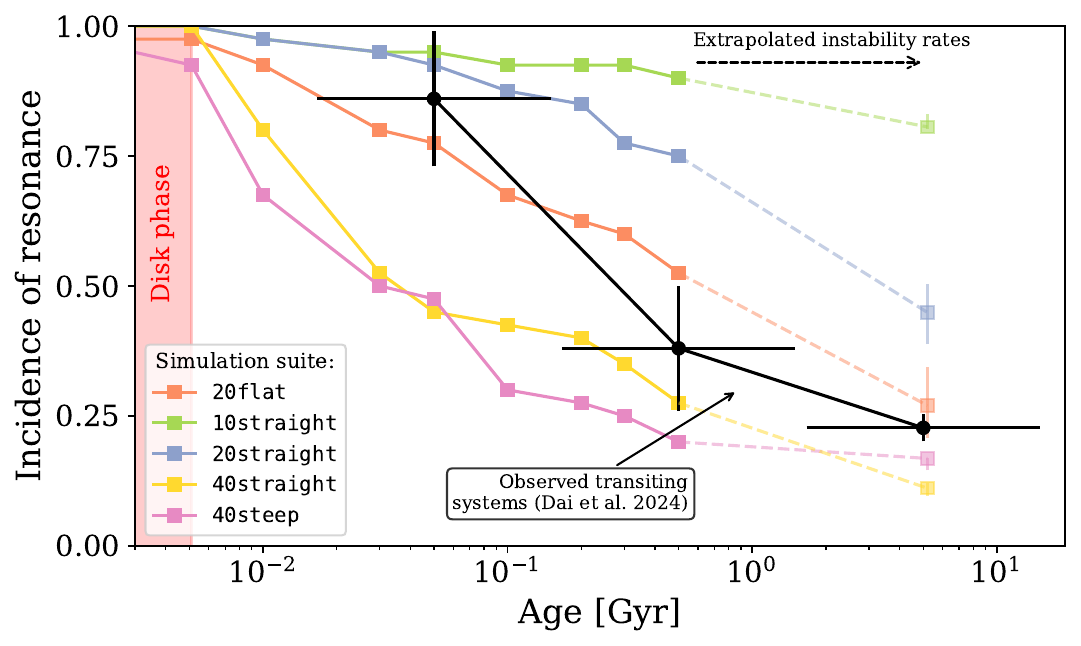}
    \caption{The fraction of simulated and observed systems as a function of age judged as ``near-resonant'' according to the empirical criterion from \cite{Dai2024}. Colored lines are each of our simulation suites, including extrapolation to 5 Gyr (see text) and the black line is the \cite{Dai2024} analysis of the observed young and field transiting systems. Nearly all systems leave the gas disk phase as resonant chains. As these chains break, fewer planet pairs are seen to be near resonance, although even non-resonant systems are sometimes determined to be near-resonance coincidentally.}
    \label{fig:resonantfrac}
\end{figure}

Figure~\ref{fig:resonantfrac} shows this empirical criterion applied to our five simulation suites as a function of time including the extrapolated instability fractions. The results follow similar trends to Fig.~\ref{fig:insttime}. The \texttt{10straight} and \texttt{20straight} simulations have too few instabilities to reproduce the $38\%$ near-resonance incidence for systems between 100~Myr and 1~Gyr. On the other hand, \texttt{40straight} and \texttt{40steep} have very early instabilities that result in few detected resonant systems at 50~Myr, in conflict with the data. The best result is obtained in the \texttt{20flat} simulations, which remain mostly resonant at 50~Myr but steadily continue breaking out of resonance through the entire integration. Furthermore, the incidence of resonance in this model extrapolated to 5~Gyr is very close to the measured value.

To summarize briefly, the \texttt{10straight} and \texttt{20straight} are a poor fit to the data, being consistently too low mass, too resonant, and too compact. In contrast, \texttt{20flat}, \texttt{40steep}, and \texttt{40straight} qualitatively reproduce many aspects of the data well. However, each of them fails at least one constraint. \texttt{20flat} has under-excited inclinations, leading to too few single-transiting systems and too compact observed systems. \texttt{40steep} and \texttt{40straight} are a better match to the \textit{Kepler} statistics but have too many early instabilities. It is likely that a mixture of these models would be an even better fit \citep{Goldberg2022}. Generically, there is a wide variety of protoplanetary disks and the resulting planetary systems inherit this diversity \citep{Emsenhuber2021a}.

\section{Discussion}
\label{sec:discussion}

\subsection{Why do some systems go unstable?}
\label{sec:unstabphysics}
\begin{figure*}
    \centering
    \includegraphics[width=0.66\linewidth]{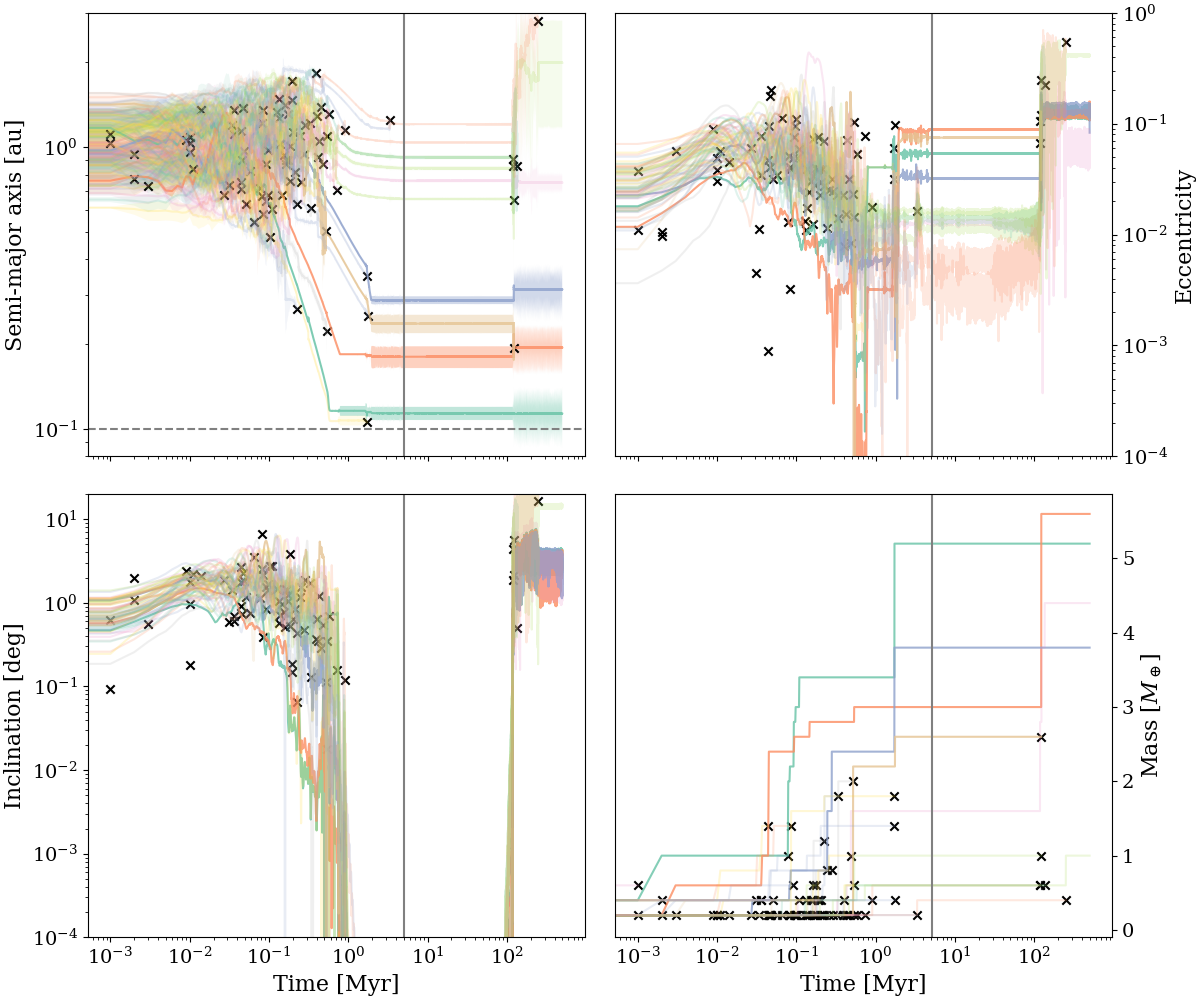}
    \includegraphics[width=0.32\linewidth]{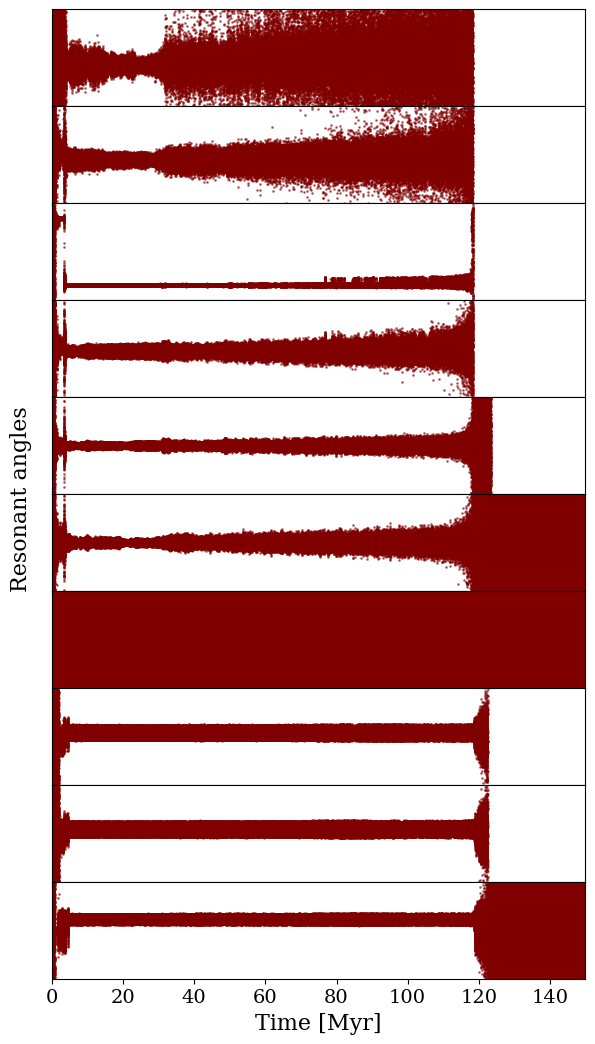}
    \caption{An example of a system from the \texttt{20flat} suite that experienced a post-disk instability. Left: evolution of the orbital elements and masses over time. Disk dissipation is marked by a vertical line at 5.1 Myr. The horizontal dashed line in the top left panel marks the planet trap at 0.1 au representing the disk inner edge. Black Xs mark when planets were removed from the simulation after a collision and merger. This system clearly has distinct outer and inner chains separated by a period ratio gap of 3.45. Right: mean motion resonant angles of adjacent pairs of planets in the system from out to in, with resonances determined from the state of the system at the time of disk removal. The y-axis in each panel ranges from 0 to $2\pi$. The x-axis is scaled linearly. All pairs except the planets across the gap are in a librating resonant state at the time of disk dissipation, but the libration amplitudes of the outer chain increase over a timescale of $\sim \rm 100~Myr$, eventually entering circulation and triggering an instability. It appears that this perturbation was passed to the inner chain, making it unstable.}
    \label{fig:unstabsystem}
\end{figure*}

Figure~\ref{fig:unstabsystem}, showing a typical example of the the \texttt{20flat} simulation suite, gives a clue as to how these initially resonant systems can slowly unravel. Upon exiting the disk phase, this system contains two resonant chains separated by a large gap of 3.45 in period ratio. The inner chain is relatively widely spaced, with pairs in the 2:1, 3:2, and 4:3 resonances. The outer chain is far more compact, containing pairs in 5:4, 7:6, 8:7, 1:1 (co-orbital), 6:5, and 5:4 resonances. All resonances are librating upon exiting the disk phase, although the outermost pairs have fairly high libration amplitude. From then until $t=\rm 30~Myr$, it is clear that pure N-body dynamics are causing the system to chaotically diffuse in action space, increasing the amplitude of libration of some pairs while decreasing that of others. Around $t=\rm 30~Myr$, the outermost resonance rapidly grows in libration amplitude and then begins sporadically circulating, which starts to gradually excite the libration of the remaining resonances in the chain. As the eccentricities grow, new resonant modes appear for the co-orbital pair (near $t=\rm 80~Myr$) and the 6:5 resonance enters circulation. At around $t=\rm 110~Myr$, the growth in libration amplitudes suddenly accelerates and the entire outer chain breaks simultaneously. Without the protection against close conjunctions provided by the resonant libration, the outer planets excite each other onto crossing orbits and collide, leaving just two survivors.

During nearly this entire process, the inner four-planet chain remains effectively unchanged. All resonances librate at constant amplitudes and chaos is not apparent. However, at the moment that the outer chain becomes unstable and its planets reach $e\gtrsim 0.1$, the libration amplitudes in the inner chain jump to a large amplitude. From there, they quickly grow until they reach circulation, the inner chain becomes unstable, and the middle two planets merge. The final system is non-resonant and excited in eccentricity and inclination.

A particularly remarkable fact is that the inner system appears to be stable on its own. While it receives extra secular precession from the outer system, the separation of timescales between secular and resonant motion ensures that it precesses as a solid object without affecting the mean motion resonances \citep{Petit2017,Tamayo2021}. The outer cluster of planets is unlikely to be observed in transit, especially after gaining significant mutual inclination during the instability, yet plays the key role in the evolution of the entire system. {We refer the interested reader to the recent work by \cite{Ogihara2026} for details on the importance of secular perturbations from an outer system to instabilities of inner resonant chains.}

\begin{figure*}
    \centering
    \includegraphics[width=0.66\linewidth]{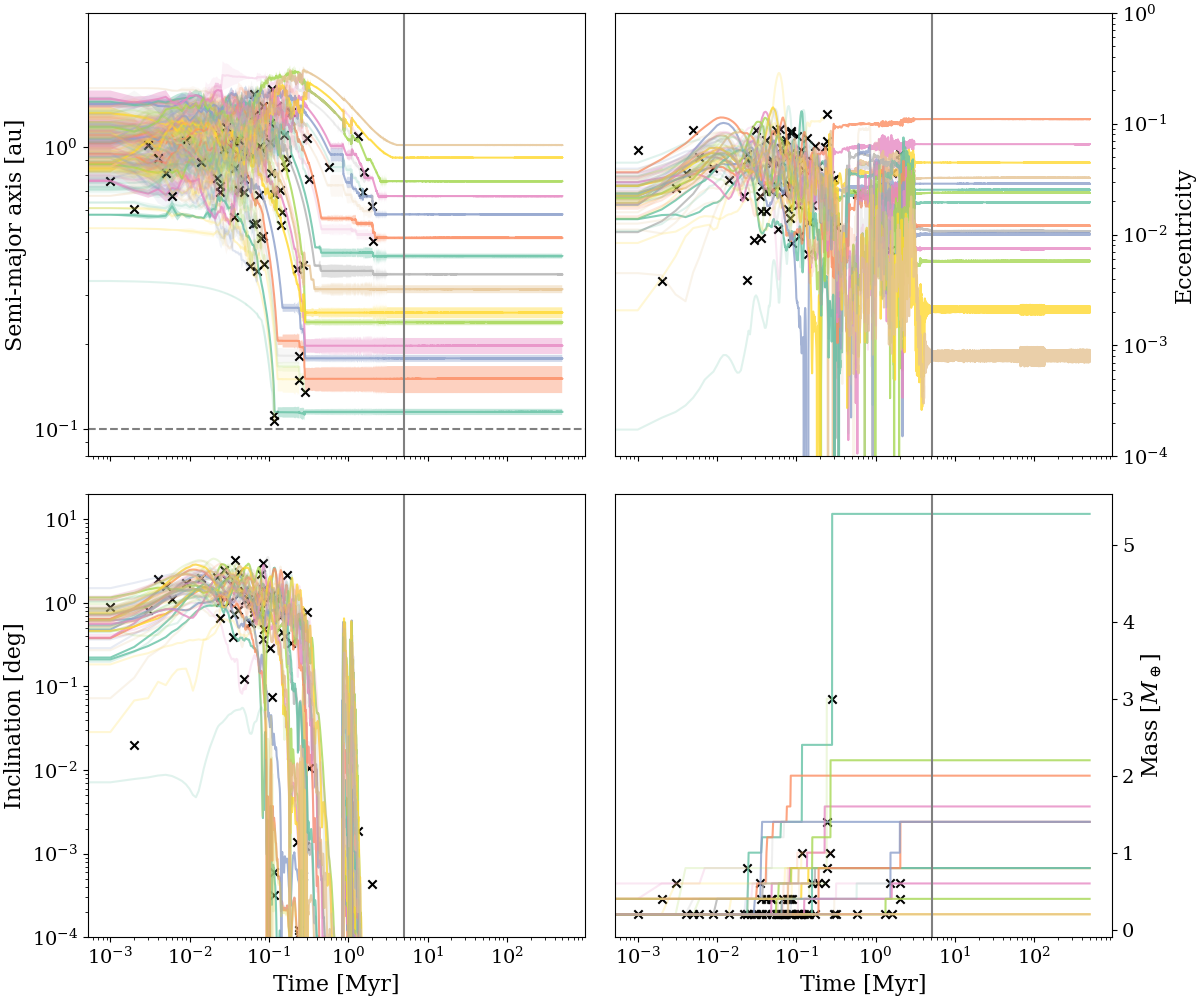}
    \includegraphics[width=0.31\linewidth]{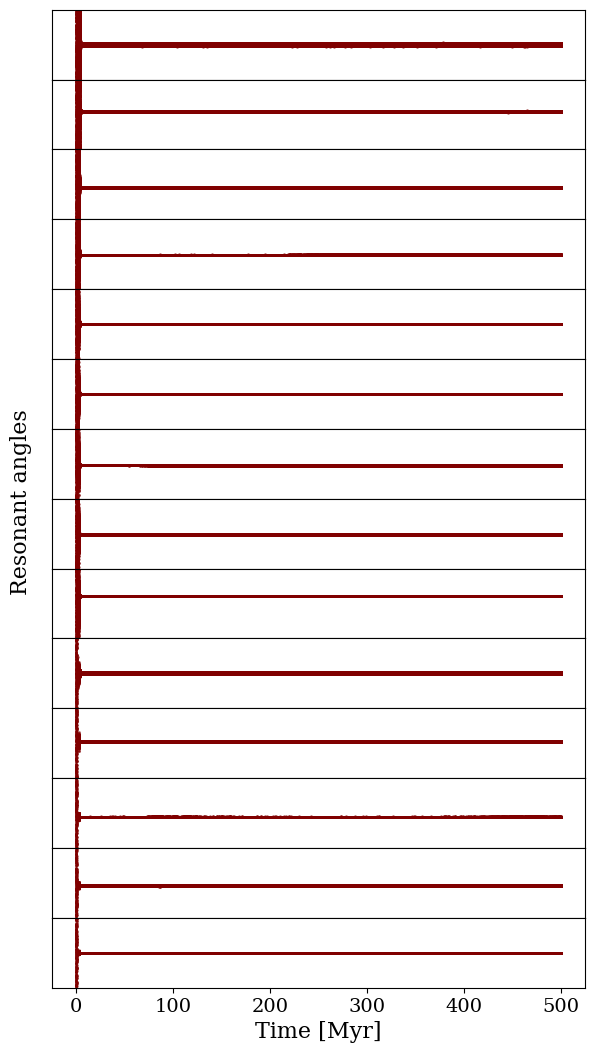}
    \caption{An example of a system from the \texttt{20straight} suite that remained stable and in resonance for the entirety of the simulation. Left: evolution of the orbital elements and masses over time. Disk dissipation is marked by a vertical dashed line at 5.1 Myr. The horizontal dashed line in the top left panel marks the planet trap at 0.1 au representing the disk inner edge. Black Xs mark when planets were removed from the simulation after a collision and merger. Right: mean motion resonant angles of adjacent pairs of planets in the system from out to in, with resonances determined from the state of the system at the time of disk removal. The y-axis in each panel ranges from 0 to $2\pi$. All pairs remain in a low-amplitude librating resonant state from the time of disk dissipation to the end of the simulation.}
    \label{fig:stabsystem}
\end{figure*}

This two-phase, outer to inner mode is the most common way that resonant chains break in our simulations, but other pathways are possible. Some outer chains go unstable but the inner chain is unaffected (these are counted as stable in our analysis). In a few cases, the instability clearly arose from the inner planets and spread to the outer system. Finally, in other cases it is ambiguous which set of planets is the main source of diffusion. 

In general, it is not clear why some resonant chains go unstable, although one instability in the three-planet case is understood \citep{Pichierri2020} and heuristics for longer chains are known \citep{Goldberg2022a}. We hypothesize that the higher incidence of instability in the simulations with a pressure bump is because low masses and slow migration in the 1 au region consistently create a very compact resonant chain. Although this outer chain is stable to chaotic overlap of adjacent first-order resonances \citep{Deck2013}, a large network of higher-order resonances can drive diffusion away from the resonant center and, once the instability is eventually triggered, an impulsive kick can be passed to the inner planets. Instabilities rarely occur in the inner chain because planets massive enough to saturate the corotation torques and reach the inner system typically capture into wider resonances which are less prone to resonant overlap and chaos. In contrast, in purely power-law disks (e.g., Fig.~\ref{fig:stabsystem}), there is no clear inner/outer system separation. The outermost planets have very low masses and the entire chain has very low resonant libration amplitude.

\subsection{Remaining resonant systems}
\label{sec:reschains}
A fundamental prediction of the breaking-the-chains model is that the surviving stable systems should closely resemble the observed resonant chains \citep{Bitsch2024}. The simplest comparison is to the orbital period architectures; we will focus on period ratios of adjacent planets and proximity to two- and three-body resonances. For each adjacent planet pair with periods $P_\textrm{in}<P_\textrm{out}$, we describe proximity to two-body resonance by the parameter
\begin{equation}
    \Delta = \frac{P_\textrm{out}}{P_\textrm{in}}\frac{p}{p+q} - 1
\end{equation}
where we consider only the closest first- or second-order $p+q$:$p$ resonance. To estimate proximity to zeroth-order three-body resonance, we follow the technique of \cite{Goldberg2021}. For each planet triplet in the system with mean motions $n_1=2\pi/P_1>n_2=2\pi/P_2>n_3=2\pi/P_3$, we compute the value
\begin{equation}
    B = p n_1 - (p+q)n_2 + qn_3
\end{equation}
for all positive coprime integer values of $p$ and $q$ with $p+q<14$ and normalize it by $\langle n\rangle = (n_1+n_2+n_3)/3$. Then, only the smallest value of $|B|/\langle n\rangle$ is kept for each system.

\begin{figure}
    \centering
    \includegraphics[width=\linewidth]{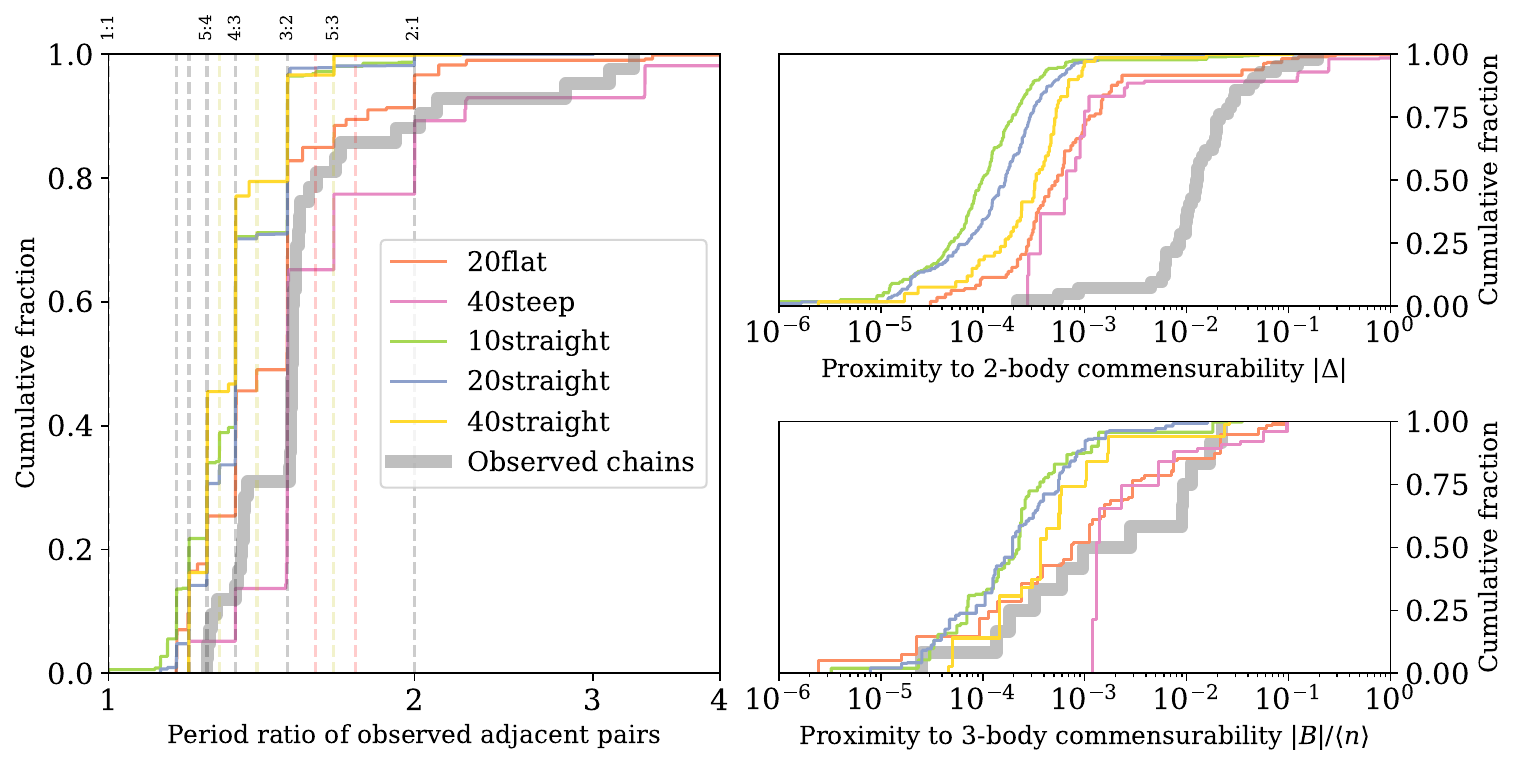}
    \caption{Period ratios of adjacent planets and proximity to two- and three-body resonances for our simulated systems that remained stable during the 500~Myr integrations. A synthetic observation filter has been applied to select for systems that would be observed in transit. In the thick gray line are observed exoplanet systems believed to be resonant chains.}
    \label{fig:obschain}
\end{figure}

Figure~\ref{fig:obschain} shows these statistics for each simulation suite after applying the synthetic observations filter from Section \ref{sec:synthobs}. We also plot these statistics for the observed 3+ planet resonant chain sample identified by \cite{Kajtazi2023} with the addition of TOI-1136 \citep{Dai2023} and again removing systems with giant planets. Simulations suites \texttt{20flat} and \texttt{40steep} closely match the observed period ratio distribution, in particular reproducing the strong abundance of 3:2 pairs. In contrast, the simulations with pure power law disks produced too many systems with 5:4 and 4:3 pairs compared to the observed population. However, our simulated planet pairs are much closer to exact resonant commensurability, as expressed by $\Delta$ (top right panel), than observed. Nevertheless, the proximity to three-body commensurability is very close the observations, especially for \texttt{20flat} and \texttt{40steep}. In other words, many systems show a $\sim1\%$ positive deviation from two-body resonance while being clearly in three-body resonance. The inability of Type I migration to reproduce this feature of the observations is well known \citep[e.g.,][and references therein]{Charalambous2023}. Tidal dissipation, which we do not include in our simulations for simplicity, can increase $\Delta$ while preserving $B$ and potentially explain this mismatch \citep{Papaloizou2015,Goldberg2021,Charalambous2023}.

Finally, we note that many of these chains have more detailed mass and eccentricity constraints measured by transit timing variations (TTVs), but a detailed comparison to these inhomogeneous analyses is left to further work.

\subsection{Neglected effects}
Our scheme neglects physics that may be important in planet formation. Real disks are turbulent and these stochastic eddies can suppress capture into resonance during migration \citep{Batygin2017,Chen2025}. The true impact will depend on the location of the magneto-rotational instability turn-off \citep{CevallosSoto2025}. Nevertheless, this choice appears to be justified. \cite{Izidoro2017} considered the effect of stochastic kicks caused by turbulence in their model with a much larger $\alpha$ than we use and found that the results were nearly indistinguishable from non-turbulent runs. 

Disks are also not static or uniformly depleting in time. Our surface density profile is inspired by the results of \cite{Suzuki2016}, who consistently found a flat or increasing surface density in the inner disk. However, the position of the peak in surface density evolves outwards in time as the disk dissipates. \cite{Ogihara2024} performed N-body simulations with Type I migration directly using these disk models and found similar super-Earth systems to ours. It is likely that the details of the disk profile are most important when planets are $\sim \rm 1~M_\oplus$ and the corotation torque can act as a mass filter. This occurs in our simulations around $\rm 100~kyr$ to $\rm 1~Myr$, when our disk profiles closely match the results of \cite{Suzuki2016}.

We also ignored interior structure and atmospheric evolution of the forming planets by always assuming their density to be $\rm 3~g~cm^{-3}$. In reality, young planets are expected to have puffy, extended envelopes that slowly cool and contract. The connection between this and widespread giant impacts after the gas disk dissipation is not clear. Although SPH simulations of super-Earth and sub-Neptune collisions imply relatively little atmospheric loss from the mechanical shock \citep{Denman2020}, \cite{Biersteker2019} proposed that post-collision heating would thermally expand the envelope and lead to complete loss via Parker winds. \cite{Izidoro2022} showed that this effect, along with expanded radii of water-rich planets that accreted outside of the ice line, can even produce a radius gap between super-Earths and sub-Neptunes. 

The radii themselves also have dynamical importance. Large planets are more likely to collide, terminating instabilities before too much dynamical excitation. But, it is not clear what the proper radius to use is, because overlap of atmospheres may not be sufficient for collision and merger \citep[see discussion in][]{Kimura2025}. We therefore chose $\rm 3~g~cm^{-3}$ as a compromise between core-determined and envelope-determined collision and merger.

\subsection{Is something missing?}
We return to the question of whether pure N-body dynamics of planetary systems are sufficient to break almost all resonant chains and generate the observed small transiting planet sample. It appears that many features of the exoplanetary census are indeed reproduced by our simplified model, at least for certain choices of parameters. Specifically, the observed period ratio and resonant fraction lie between our \texttt{20flat} and \texttt{40steep} models, suggesting that a range of available solids to form rocky planets is necessary to match the data. 

Interestingly, the strength of the observed \textit{Kepler} dichotomy is not reached even in our highest mass simulations. In other words, violent instabilities that produce enough single transiting planets also tend to make too widely-spaced systems (e.g., Fig~\ref{fig:40steepobspopstats}) and vice versa (Fig~\ref{fig:20flatobspopstats}). Our \texttt{40straight} suite was a good match to both the observed multiplicity and period ratio distribution (Fig~\ref{fig:40straightobspopstats}) but predominately makes planets of mass $\rm\sim 10~ M_\oplus$ (Fig~\ref{fig:40straightpopstats}), seemingly too high compared to the typical \textit{Kepler} super-Earth or sub-Neptune of $\rm\sim 3~M_\oplus$ \citep{He2020,Rogers2021}. This problem also appears in \cite{Izidoro2021a}, who tested multiple pebble accretion models with different mass flux. Their sole model that matches the observed multiplicity distribution (Model-I, $t_\textrm{start}=\rm 0.5~Myr$ in their notation), has a typical planet mass of $\rm 10-20~M_\oplus$.

One potential way to bridge this gap is an additional source of inclination excitation. \cite{Spalding2016} showed that mutual inclinations can be excited in multiplanet systems through the quadrupolar precession of a primordially misaligned host star. During spindown, a secular resonance can be excited that tilts the outer planet to a non-transiting configuration. Although this process can itself trigger a dynamical instability \citep{Spalding2018}, the quiescent version appears a promising mechanism to raise mutual inclinations in some systems without requiring large masses or orbital spacings between planets.

Orbital eccentricities are more difficult to compare to models because they are not directly measurable from transit observations. Transit timing variations provide more information on the eccentricity through the mean-motion resonant dynamics. \cite{Yee2021} showed that TTV systems have eccentricities several times lower than that required for stability, suggesting some amount of dissipation that damped eccentricity following a giant impact phase. However, TTV systems represent a biased sample of near-resonant planets, which, in the context of the breaking-the-chains model, are less likely to have undergone an instability that excites eccentricity \citep{Goldberg2022}.

Transit duration measurements, on the other hand, can constrain the average eccentricity of an entire population of planets and have indicated that \textit{Kepler} multiplanet systems have typical eccentricities of $\sim 0.05$ \citep{VanEylen2019,Mills2019}. \cite{Gilbert2025} refined this strategy by dividing the population in planet radius bins and found that small planets ($\rm\lesssim 4~R_\oplus$) have mean eccentricity of 0.06 in single transiting systems and 0.03 in multitransiting systems. After applying the synthetic observational bias, we find mean eccentricities of 0.06 and 0.05 in single and multitransiting systems in the \texttt{20flat} simulations and 0.09 and 0.07 in the \texttt{40steep} simulations. This is consistent with a small degree of eccentricity damping over the lifetime of the system after a dynamical instability, such as that provided by tidal dissipation \citep{Delisle2014a}.

This model predicts that the youngest observed systems should be predominantly chains of planetary deep in resonance. Although \cite{Dai2024} identified a surplus of systems near resonance according to an orbital period based criterion, most of these systems do not have sufficient dynamical constraints to determine whether they are actually resonant according to a dynamical definition that requires libration of the resonant angle within a separatrix-bounded region. \cite{Hu2025} studied several of these systems in detail and found several planet pairs that, while classified as near-resonant in terms of period ratio, are sufficiently distant from exact commensurability that their resonant angles are likely to be circulating. As near-resonance in our synthetic systems typically implies resonant libration, these intriguing observational findings imply additional physics not captured in our simulations during the disk phase or shortly after disk dissipation that can modify the resonant state.

Finally, we note that the very steep decline in the observed incidence of resonance between the first and second bin of Fig~\ref{fig:resonantfrac} is not well reproduced in our simulations. While \cite{Dai2024} presented a compelling argument for a population trend, there is reason to believe that the quantitative details are not set in stone. For one, \cite{Dai2024} did not attempt to debias the sample. Young transiting planets present numerous difficulties in detection and confirmation due to stellar activity and inhomogeneous distribution on the sky. Furthermore, near-resonant planets experience transit timing variations that smear transits after phase-folding and impede detection \citep{Leleu2021a}. Neither of these effects nor their combination has been carefully treated in the literature. We expect that the PLATO mission, which is intended to provide a robust statistical sample as well as accurately measured ages for many planet hosts, will greatly refine the observational constraints and potentially clarify whether there is in fact a preferred time of instability.

\section{Conclusion}
\label{sec:conclusion}
In this work, we performed a large suite of very long duration N-body integrations incorporating planet growth in a ring, Type I migration, and dynamical instability. Planetary systems are almost entirely resonant chains at the end of the disk phase, which gradually break over the course of our 500 Myr simulations. We compared the final systems to observed transiting planet data after applying a synthetic observation filter. 

We present the following conclusions:
\begin{itemize}
    \item We found that certain combinations of disk surface density profiles and initial solid mass can reproduce the \textit{Kepler} sample demographics as well as the observed multiplanet system resonant fraction as a function of age. This scenario requires that 90--95\% of initially resonant systems undergo an instability. The remaining resonant systems closely resemble known resonant chains in orbital architecture with deviations being explained by tidal dissipation.
    \item Pressure bumps and their resulting migration traps are effective ways to increase the rate of instabilities compared to purely power-law disks. There is frequently an outer resonant chain that is unstable on long timescales and passes its instability to the inner chain, which would have been stable on its own. These outer companions are not easily observable with current transiting or radial velocity techniques, but may be detectable by microlensing. The role of eccentric outer planets in triggering instabilities in an inner chain has also been studied simultaneously and independently by \cite{Ogihara2026}.
    \item Compact planetary systems continuously evolve over hundreds of Myr, or billions of orbits. Comparing planet formation models that end at 100~Myr or less to mature systems with ages 1--10 Gyr can be deeply misleading. The common issue that many population synthesis models produce too many resonant systems may simply be an artifact of ending N-body integrations too early. If the high fraction of mean-motion resonance in young systems reported by \cite{Dai2024} is confirmed by further observations, this may be a correct prediction of these models.
\end{itemize}

Our model is highly simplified. We intentionally neglect complex physics in order to highlight the celestial mechanics of multiplanet systems. While many population synthesis models have implemented additional physical effects such as disk evolution, dust drift and pebble accretion, or atmospheric evolution, these processes are highly uncertain. We wish to emphasize that better-understood mechanisms like planet migration, and especially Newtonian gravity, should be tested to their limits to determine where they fall short. More complex models cannot neglect dynamical evolution when comparing to known systems.

As a whole, our findings strongly support the ``breaking-the-chains'' model as the dominant pathway for the formation of systems of small planets within 100 d. It has become increasingly clear that the architectures of planetary systems are not set at the moment of disk dissipation \citep{Berger2020,Yang2023}. Current and upcoming observations with \textit{TESS} and PLATO, especially those of young systems and with well-measured host star ages, will help us map the evolution of planetary systems over their lifetimes. The onus then lies on theorists to produce models of dynamical evolution extending beyond the first 1--10\% of the lifetimes of typical systems that can be used to reliably infer details of their initial conditions and formation. This will be a formidable challenge for traditional N-body methods, but continuing advancements in accurate semi-analytical models may be a promising path forward \citep{Kimura2025,Kokubo2025}.

\begin{acknowledgements}
We are grateful to referee Dan Tamayo for useful comments that improved the paper. We thank Alessandro Morbidelli, Masahiro Ogihara, Masanobu Kunitomo, and Konstantin Batygin for insightful discussions. This work was funded by the ERC project N.~101019380 ``HolyEarth.''. A.~P. is supported by the French government, through the $ \mathrm{UCA^{J.E.D.I.}} $ Investments in the Future project managed by the National Research Agency (ANR) with the reference number ANR-15-IDEX-01.
\end{acknowledgements}

\bibliographystyle{aa}
\bibliography{PlanetFormation}

\begin{appendix}
\section{Intrinsic system statistics}
\label{app:pop}
\begin{figure}[h]
    \centering
    \includegraphics[width=\linewidth]{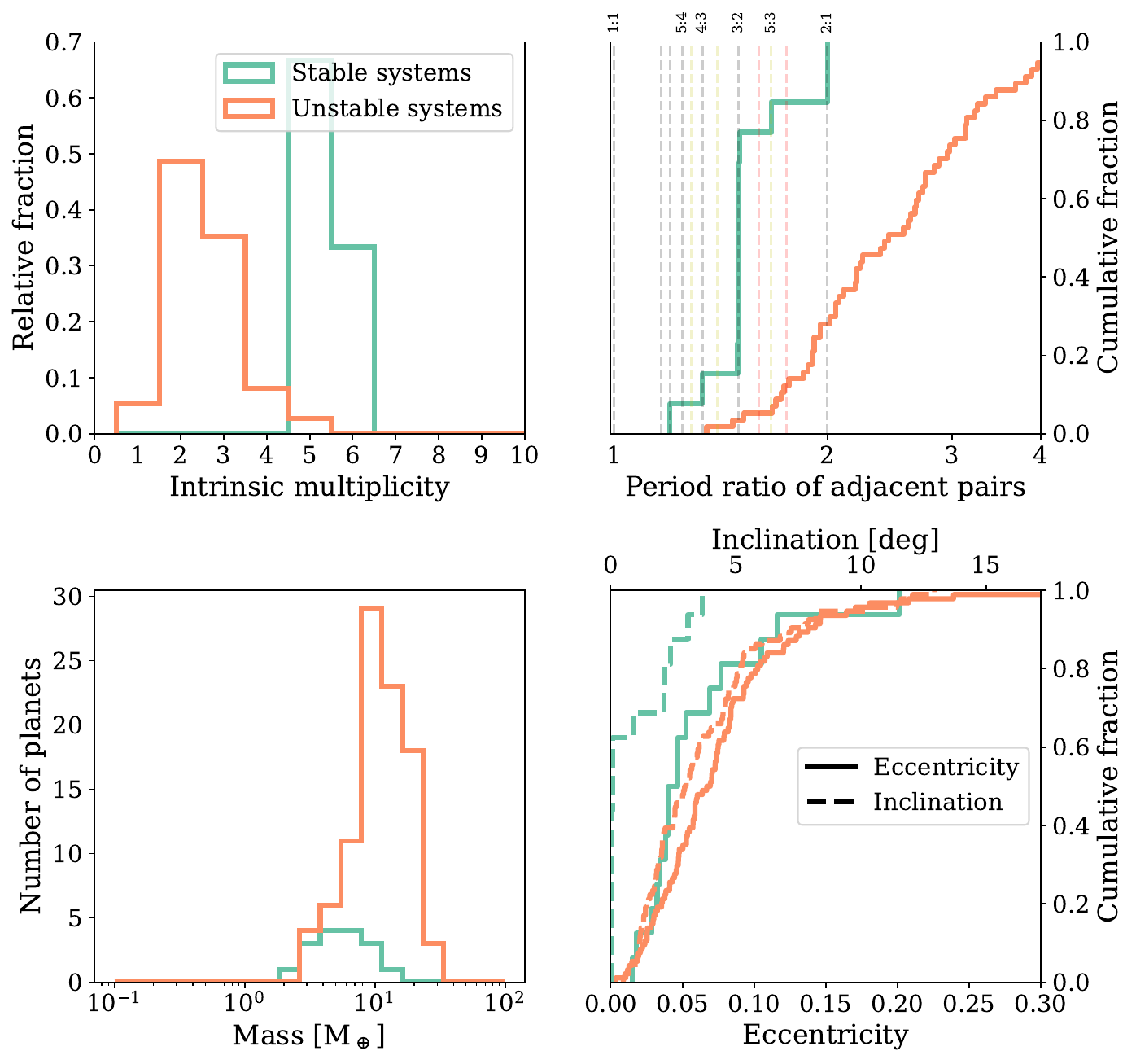}
    \caption{Same as Fig.~\ref{fig:20flatpopstats} for run \texttt{40steep}.}
    \label{fig:40steeppopstats}
\end{figure}

\begin{figure}[h]
    \centering
    \includegraphics[width=\linewidth]{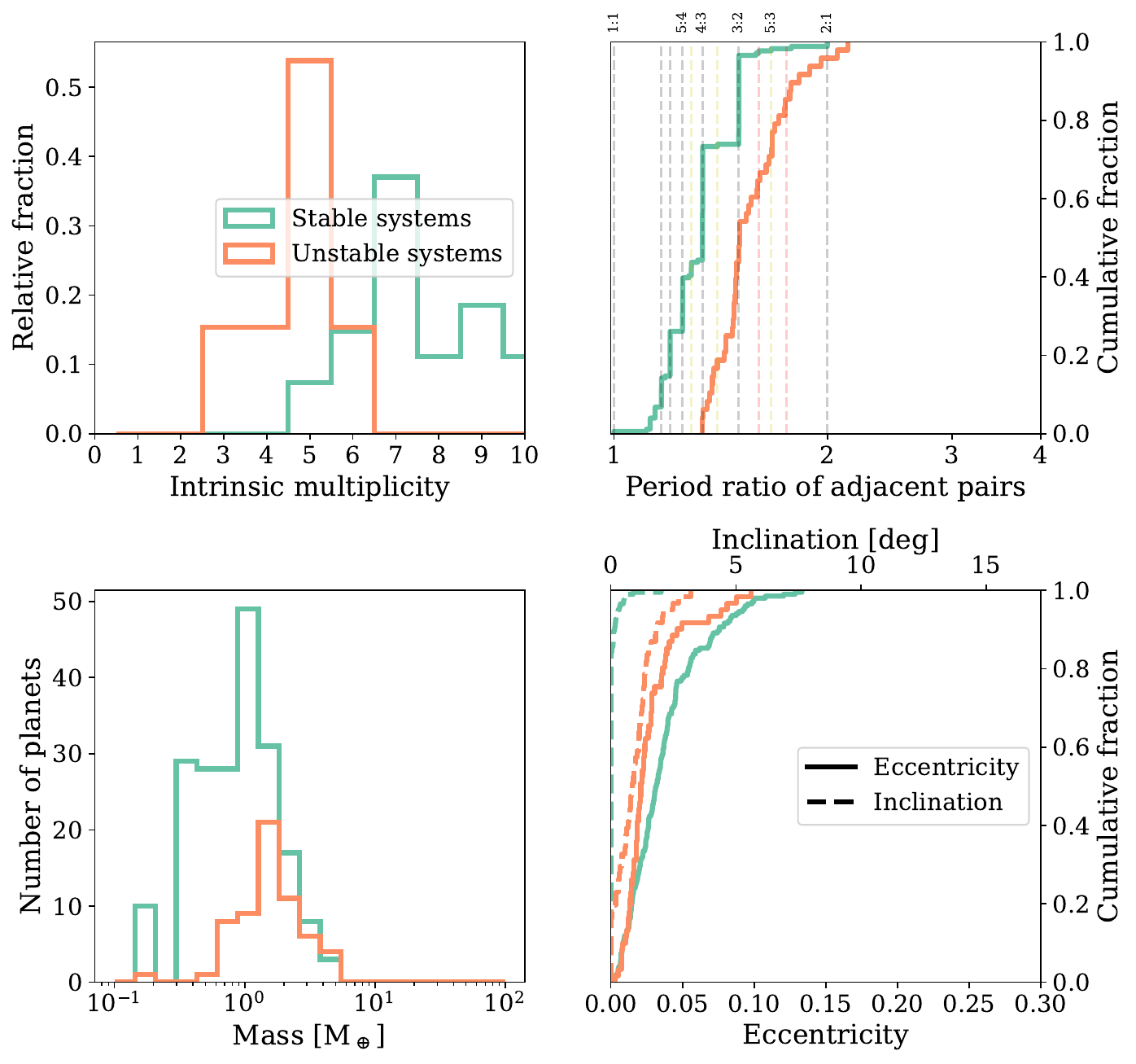}
    \caption{Same as Fig.~\ref{fig:20flatpopstats} for run \texttt{10straight}.}
    \label{fig:10straightpopstats}
\end{figure}

\begin{figure}[h]
    \centering
    \includegraphics[width=\linewidth]{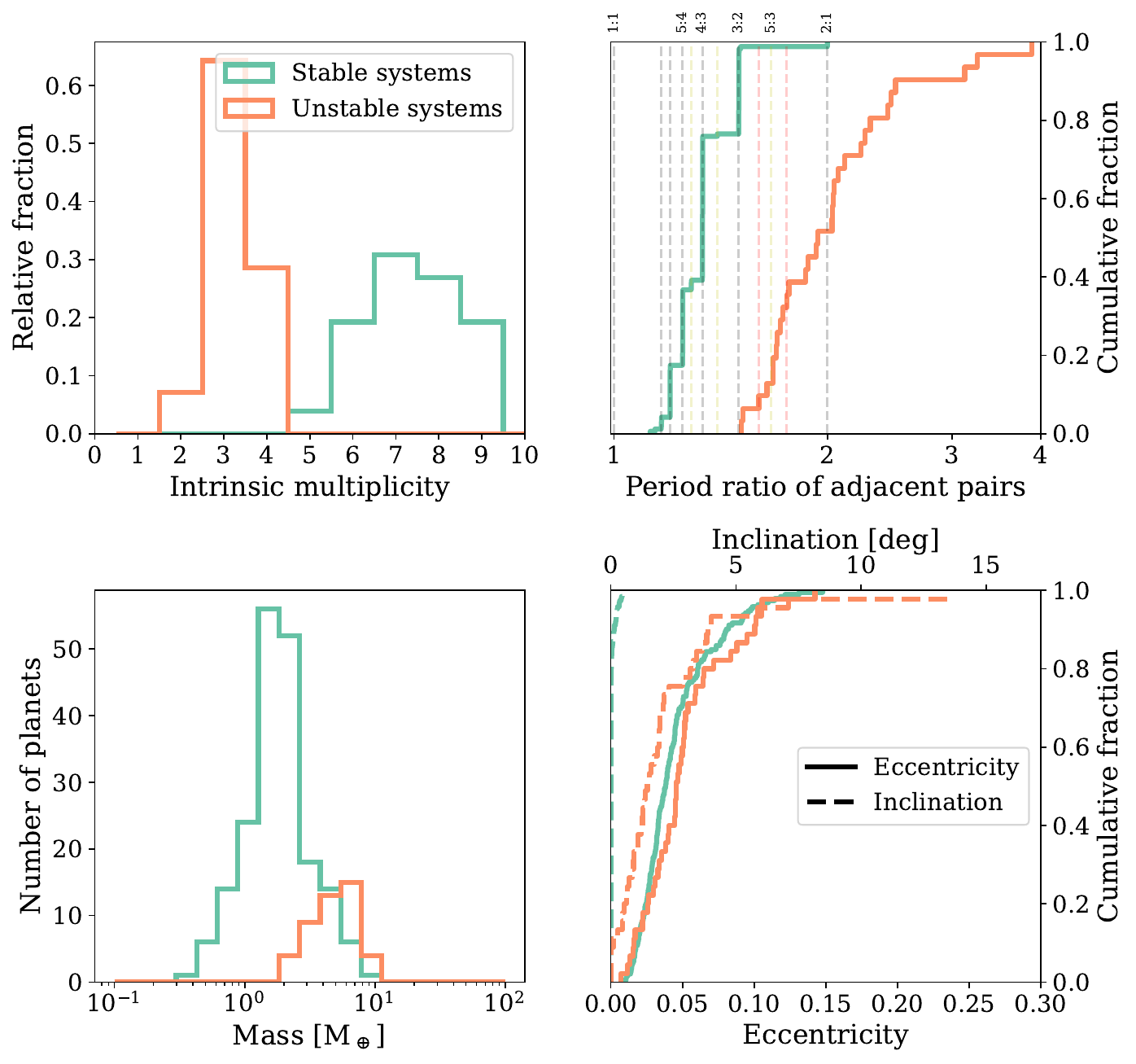}
    \caption{Same as Fig.~\ref{fig:20flatpopstats} for run \texttt{20straight}.}
    \label{fig:20straightpopstats}
\end{figure}

\begin{figure}[h]
    \centering
    \includegraphics[width=\linewidth]{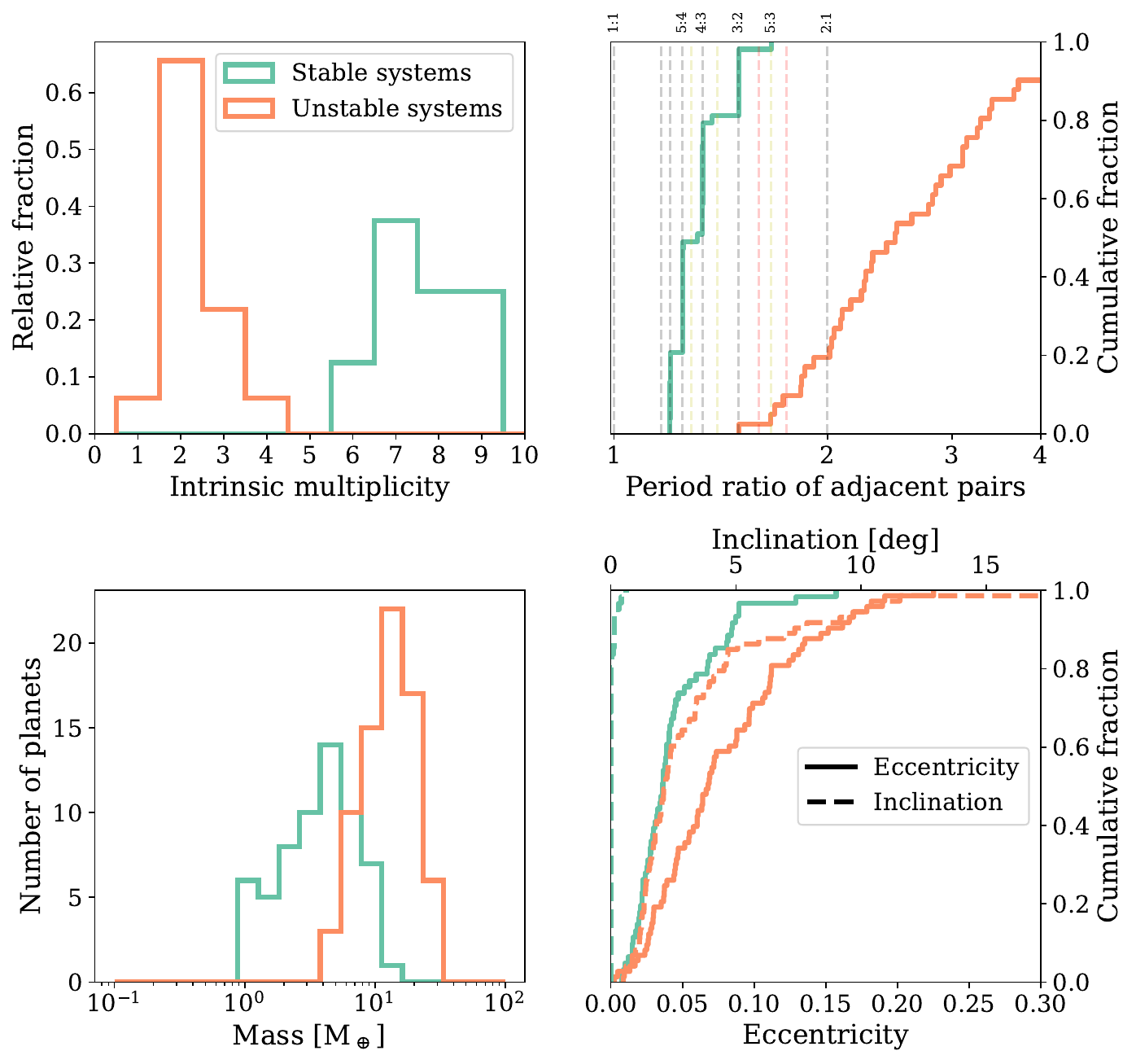}
    \caption{Same as Fig.~\ref{fig:20flatpopstats} for run \texttt{40straight}.}
    \label{fig:40straightpopstats}
\end{figure}

Figures~\ref{fig:40steeppopstats}--\ref{fig:40straightpopstats} show population statistics for the \texttt{40steep}, \texttt{10straight}, \texttt{20straight}, and \texttt{40straight} runs.

\section{Numerical tests}
We performed a series of numerical tests to ensure that the instability mechanism and timescale were not numerical artifacts caused by our choice of N-body integrator. We chose a system from the \texttt{20flat} sample that underwent an instability in the original simulation after approximately 110 Myr. This particular system had a dynamically separated inner and outer resonant chain, and the instability arose after a gradual increase in the libration amplitudes of the resonances in the outer chain. To test the robustness of this instability and the quantitative accuracy of the timescale, we removed all of the planets in the inner chain and re-integrated the outer system starting from the end of disk dissipation ($t=\rm 5.1~Myr$).

\begin{figure}[h]
    \centering
    \includegraphics[width=\linewidth]{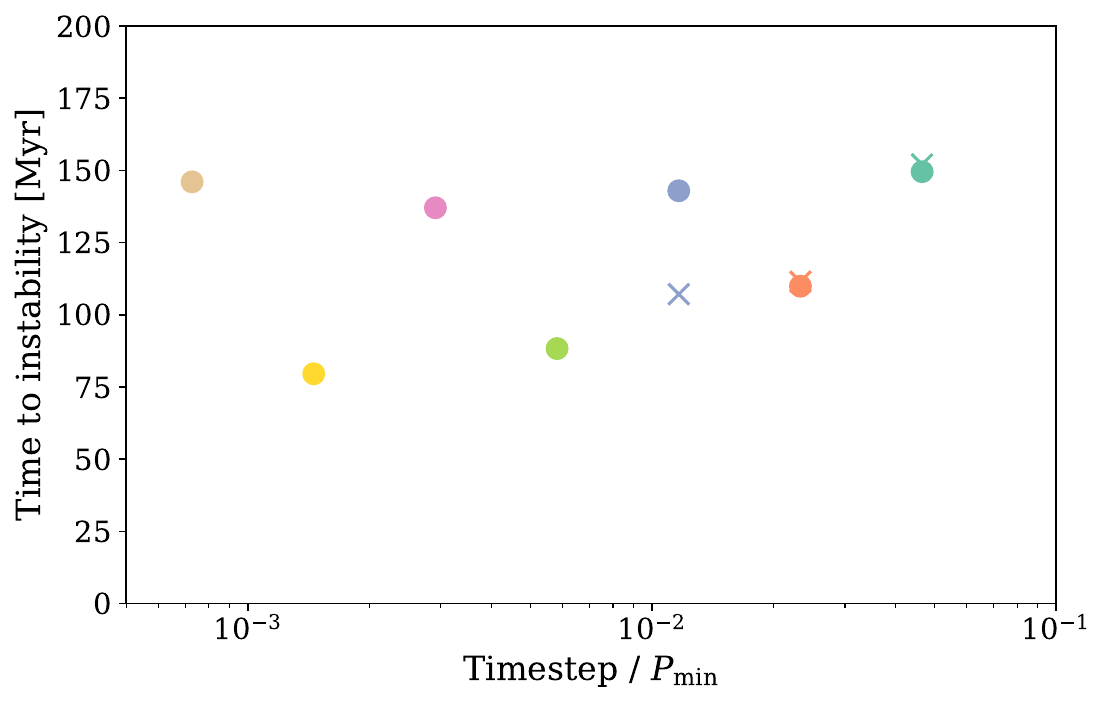}
    \caption{The time-to-instability for the outer resonant chain in one simulation from the \texttt{20flat} suite, reintegrated with different timesteps. The largest timestep is 1/20 times the period of the innermost planet. Colored circles were reintegrated with \texttt{TRACE}, which uses the Wisdom-Holman splitting, and colored crosses were reintegrated with SABA.}
    \label{fig:timestepinsttime}
\end{figure}

First, our re-integrations varied the integrator timestep as $dt_0/2^i$, where $i$ ranged from $0,1,...,6$ and $dt_0$ is 1/20 of the period of the innermost planet. Figure~\ref{fig:timestepinsttime} shows time to instability in these runs (solid circles). In each case, an instability of the same nature arose in the system, with the time to instability within a factor of two of the original value of 110 Myr. Such scatter represents a fundamental limitation in instability time predictions for chaotic systems \citep{Hussain2020}. The absence of a trend, especially the lack of an increase in the instability timescale for small timesteps, suggests that the slow growth in libration amplitude is not caused by truncation of the N-body Hamiltonian. If this effect were purely due to roundoff error or truncation error, there would be a clear increasing or decreasing trend with the timestep, respectively.

\begin{figure}
    \centering
    \includegraphics[width=\linewidth]{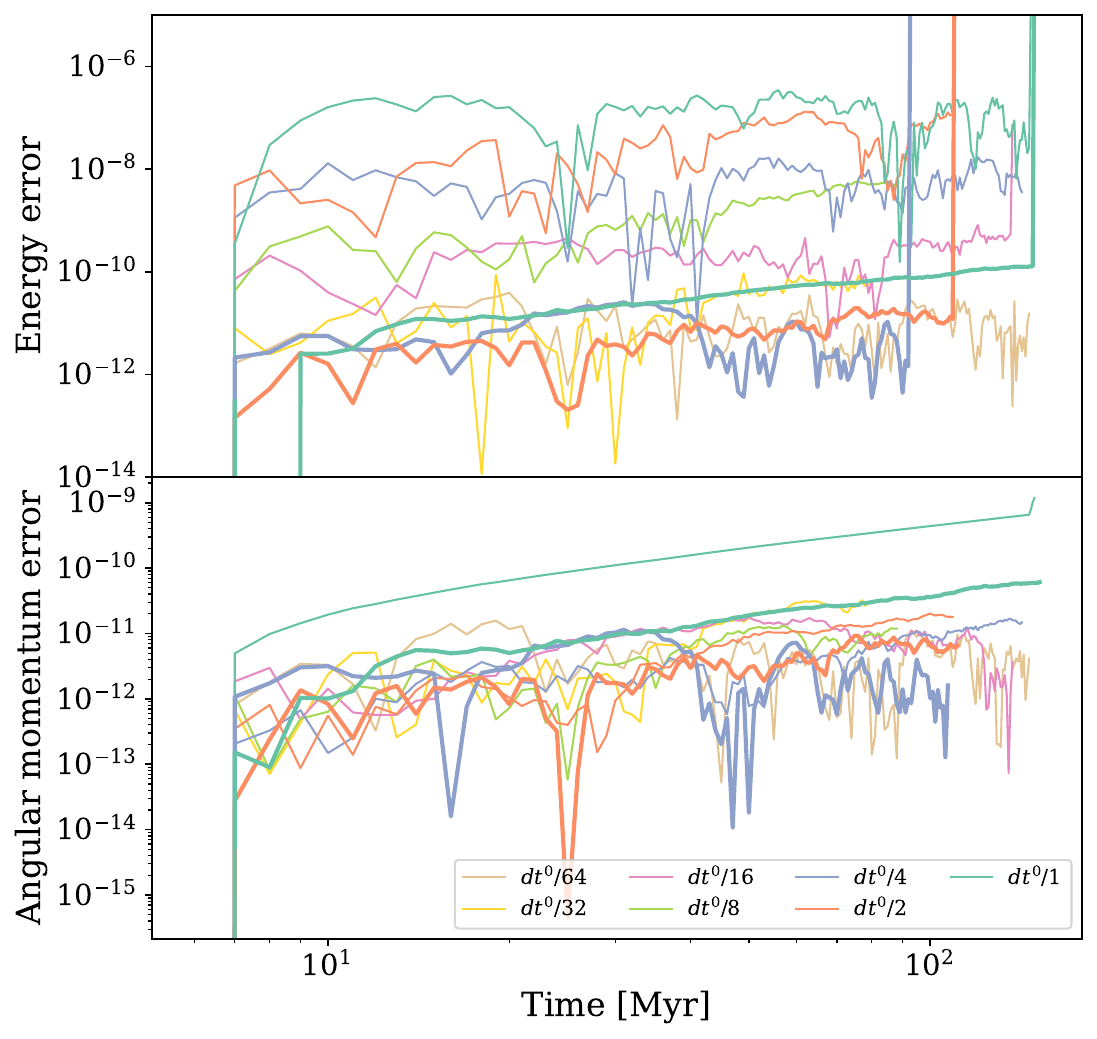}
    \caption{Energy and angular momentum error in the reintegrations. The colors are the same as Fig.~\ref{fig:timestepinsttime}. Thin lines were reintegrated with \texttt{TRACE} and thick lines with SABA.}
    \label{fig:energyerror}
\end{figure}

Second, we re-integrated the outer system with the SABA(10,6,4) symplectic integrator implemented in \texttt{rebound} \citep{Laskar2001,Rein2019a}, which has a much higher order than WHFast. SABA indeed conserves angular momentum and energy much better than WHFast for the same timestep (thick lines in Fig.~\ref{fig:energyerror}), at least until a close encounter that it was not designed to handle. However, the actual time of instability is nearly the same with SABA (crosses in Fig.~\ref{fig:timestepinsttime}) and \texttt{TRACE} with WHFast (circles). This suggests that that the instability is robust to choice of integrators and likely not a numerical artifact.

\end{appendix}
\end{document}